# Dynamic Structure-Soil-Structure Interaction for Nuclear Power Plants

Constantinos Kanellopoulos[1], Peter Rangelow[2], Boris Jeremic[3], Ioannis Anastasopoulos[1], Bozidar Stojadinovic[1]

[1]Department of Civil, Environmental, and Geomatic Engineering, ETH Zurich, Zurich 8093, Switzerland
[2]Section of Seismic Engineering & Dynamics, Basler & Hofmann AG Consulting Engineers, Zurich 8032, Switzerland
[3]Department of Civil and Environmental Engineering, University of California Davis, Davis, 95616, USA

## Abstract

The paper explores the linear and nonlinear dynamic interaction between the reactor and the auxiliary buildings of a Nuclear Power Plant on a realistic layered soil profile, aiming to evaluate the effect of the auxiliary building on the seismic response of crucial components inside the reactor building. Based on realistic geometrical assumptions, high-fidelity 3D finite element (FE) models of increasing sophistication are created in the Real-ESSI Simulator. Starting with elastic soil conditions and assuming tied soil–foundation interfaces, it is shown that the rocking vibration mode of the soil–reactor building system is amplified by the presence of the auxiliary building through a detrimental out-of-phase rotational interaction mechanism. Adding nonlinear interfaces, which allow for soil–foundation detachment during seismic shaking, introduces higher excitation frequencies (above 10 Hz) in the foundation of the reactor building, leading to amplification effects in the resonant vibration response of the biological shield wall (incl. reactor vessel) inside the reactor building. A small amount of sliding at the soil–foundation interface of the auxiliary building slightly decreases its response, thus reducing its aforementioned negative effects on the reactor building. When soil nonlinearity is accounted for, the rocking vibration mode of the soil–reactor building system almost vanishes, thanks to the strongly nonlinear response of the underlying soil. This leads to a beneficial out-of-phase horizontal interaction mechanism between the two buildings, reducing the spectral accelerations at critical points inside the reactor building by up to 55% for frequencies close to the resonant vibration frequency of the auxiliary building. This implies that the neighboring buildings could offer mutual seismic protection to each other, in a similar way to the recently emerged seismic resonant metamaterials, provided that they are properly tuned during the design phase, accounting for soil and soil-foundation interface nonlinearities.

# 1 Introduction

Following the recent advancements in computing technology, and the development of parallel and high-performance computing (HPC) infrastructures that allow faster and more complex numerical simulations, a considerable amount of research has been devoted to the development of more realistic numerical models for soil-structure interaction (SSI) problems. Thanks to the enhanced computing power, it is yet possible to account for computationally demanding system nonlinearities, including the soil, the structure, and the various interfaces. The importance of accounting for SSI and for such nonlinearities in earthquake engineering models is well-known and is supported by numerous studies [1–6]. However, the research on Structure-Soil-Structure Interaction (SSSI) is still at an early stage, and its effects on the nonlinear dynamic interaction between important structures, such as Nuclear Power Plants (NPPs), are yet to be explored.

The first study on SSSI was by MacCalden back in 1969 [7], who investigated both analytically and experimentally the transmission of steady-state vibrations from one rigid circular foundation to another through the soil. Two years later, Warburton et al. [8] studied the response of two cylindrical masses on an elastic half-space, excited by a harmonic force applied to one of the masses. They showed that the interaction between the masses is maximized at frequencies associated with their resonances. In 1973, Luco & Contesse [9] were the first to introduce the term SSSI, studying the dynamic interaction between two neighboring structures, modeled as shear walls on rigid semi-circular foundations, excited by vertically propagating in-plane SH waves. Important interaction effects were observed in the case of a small shear wall located close to a larger one, with the response of the small structure being significantly affected by the presence of the larger structure. For more realistic 3D models, coupling between the horizontal, vertical, rocking, and torsional motions of the foundations should be expected. Since then, SSSI has attracted the interest of many researchers [10–18]. Gonzales [10], and Roesset et al. [11] both concluded that the interaction between structures results in coupling between different vibration modes (e.g., vertical vibrations under horizontal force) that would not appear for a single structure. They also explained that the additional inertia of one of the structures will lower the resonant frequency of the other. Betti [14] studied the dynamic interaction between embedded foundations, showing that in the lower frequency range, the translational, rocking, and torsional components of the impedance matrix exhibit clear effects of cross-interaction or coupling (up to 30%), which decreases with increasing frequency and distance between the adjacent foundations. Padron et al. [16] studied numerically groups of structures and found that SSSI effects are important for structures of similar dynamic characteristics, resulting in either amplification or attenuation of the response. Similar conclusions have been drawn by Liang et al. [17,18], who explored 2D dynamic SSSI for twin buildings in layered half-space, for incident SH- and SV- waves. Recent experimental studies [19,20] provide further evidence of the significant role of SSSI in the dynamic response of adjacent structures, being either beneficial or detrimental. Thus, neglecting the interaction between two adjacent structures may not be conservative.

All of the aforementioned numerical studies assume linear soil and structure response and tied soil-foundation interfaces. There are only a few numerical studies that have attempted to account for nonlinear material behavior. Bolisetti & Whittaker [21] investigated the influence of SSSI in low- to mid-rise buildings including soil and interface nonlinearities. Although SSSI was found to be negligible in their studied cases, they concluded that future studies should further investigate the effect of nonlinear interfaces on SSSI problems. Long et al. [22] also emphasized the fact that existing numerical studies have insufficiently considered nonlinearities in their recent 2D numerical study on SSSI effects for high-rise buildings, in which they employed nonlinear models for soil, buildings, and interfaces. More recently, Kassas et al. [23] studied the effects of SSSI on the seismic response of neighbouring structures on liquefiable soil, concluding that such interaction may have a major effect on the accumulation of foundation rotations and settlements.

Having reliable models, capable of realistically reproducing the seismic response of soil-structure systems is of utmost importance, especially for critical infrastructure such as NPPs, where failure could have catastrophic consequences. As early as 1972, Lee & Wesley [24] suggested that NPPs can be designed to achieve a reduction in seismic loads by taking advantage of the interaction between neighboring structures, while improper design and layout could result in double resonance phenomena, and to increase of the seismic loads. Matthees & Magiera [25] and Imamura et al. [26] also found the interaction between such massive structures to be of importance, suggesting that it cannot be disregarded. These early findings are further supported by the recent work of Roy et al. [27], who examined the impact of SSSI on the In-Structure Response Spectra (ISRS) of a light nuclear structure adjacent to a heavy one. Performing linear elastic SSI analyses of the light structure using the motions adjacent to the heavier structure as input, they showed that the peak ISRS increased by a factor of even more than 3.5, underlining the significance of SSSI. Therefore, the realistic evaluation of SSSI effects in the seismic response of new and existing NPPs is rather essential, calling for advanced numerical simulations, accounting for system nonlinearities. Although such nonlinear SSI effects have been studied for a single nuclear reactor building in the past few years [28–31], to the best of our knowledge, nonlinear SSSI effects on the seismic response of NPPs are yet to be addressed.

The studies discussed above have clearly demonstrated the importance of considering nonlinear SSSI effects for adjacent structures in NPPs. Aiming to contribute towards bridging the existing gap in the literature, this paper develops a high-fidelity 3D Finite Element (FE) model of an idealized NPP, consisting of the main reactor building and an auxiliary building that surrounds it. The scope of the present study is twofold. First, the response of the reactor building with or without the presence of the auxiliary building is explored, progressively increasing the level of FE analysis sophistication. A linear elastic analysis is initially conducted, followed by the introduction of nonlinear soil-foundation interfaces, and finally of nonlinear soil response. In this way, it is possible to explore the role of each nonlinearity and to quantify its role on SSSI. Second, the effect of the auxiliary building on the seismic response of the reactor building is explored for all levels of FE analysis sophistication, giving emphasis on critical internal components, such as the reactor vessel.

# 2 Methodology

To investigate the dynamic interaction between the NPP reactor building and the auxiliary building, a detailed 3D FE model is implemented in the Real-ESSI Simulator [32] (http://real-essi.info). Real-ESSI provides state-of-the-art tools for computational modeling in earthquake engineering, enabling the development of realistic models of earthquakes, soils, structures, and their interactions (ESSI). Such tools include the Domain Reduction Method (DRM), which allows simulation of all types of seismic waves (as explained later in more detail); a variety of advanced constitutive models to simulate the nonlinear response of soils, structures, and their interaction; analytic probabilistic modeling tools; analytical calculation of all energy variables to monitor how the energy is distributed in the model; and parallel computing capabilities.

**Figure 1** shows an overview and a cross-sectional view of the 3D FE NPP–soil model, portraying the NPP reactor building positioned at the center, surrounded by the auxiliary building. The two buildings are not in direct contact with each other, having a clearance of 1 m. In reality, the clearance between such structures is significantly smaller (typically 0.1 to 0.2 m). However, in order to ensure a computationally stable simulation with a reasonable element size, a 1 m clearance is assumed. The FE mesh is generated using the open-source mesh generator Gmsh [33]. All simulations were performed on the Euler parallel computing cluster, operated by the ETH Zurich High-Performance Computing group. The results are visualized with ParaView (https://www.paraview.org/). A detailed description and verification of individual model components is offered below.

## 2.1 Domain Reduction Method (DRM)

The Domain Reduction Method is a simple, yet powerful numerical technique that allows realistic simulation of all seismic (body and surface) waves in a reduced domain, comprising only the structure(s) and soil layers of interest. Originally developed by Bielak et al. [34,35], it aims at bridging the gap between seismologists and earthquake/civil engineers. Seismologists develop regional simulation models of earthquakes, focusing on modelling the fault rupture and the consequent wave propagation through the earth's crust, without considering the structure(s). In contrast, earthquake engineers study a much smaller domain, containing only the structure(s) and the local soil of interest, (over)simplifying the input seismic waves to 1-, 2- or 3-component motions, by shaking horizontally and vertically the base of their models. Such engineering approach may produce the desired acceleration time histories at the ground surface, but it assumes that all ground surface points move identically, which may not be realistic depending on the problem being studied. With the DRM, it is possible to input arbitrary 3D seismic waves generated by the regional simulation model into the reduced domain model of interest (which can also be nonlinear), by applying equivalent effective seismic forces on the elastic DRM elements (**Figure 1**), calculated according to the DRM theory. To attenuate the radiated waves resulting from the vibration of the structures, a few layers of elastic damping elements with progressively increasing damping properties are required to surround the reduced domain model and the DRM elements. If no structures are modeled (free–field), no waves are expected to leave the reduced domain model, resulting in zero displacements of the damping elements.

Since regional earthquake model data is limited due to its high computational cost and required knowledge of geology at large depths, this study utilizes DRM to generate a vertically propagating 1-component wave field of horizontal shear waves (SH). For this purpose, Real-ESSI assumes that the regional earthquake model is a 1D wave propagation model in a layered half-space, which can be solved analytically to calculate the effective DRM forces. In other words, the only input required by the software is the acceleration time history at the desired soil depth. In this study, an artificial ground motion is generated to match the "ENSI–2015" [36] mean uniform hazard spectrum, with an annual frequency of exceedance of $10^{-4}$, targeted at the ground surface using the DRM (**Figure 1b**). Five rows of damping elements with increasing Rayleigh damping from 0% to 100% are modeled around the reduced domain.

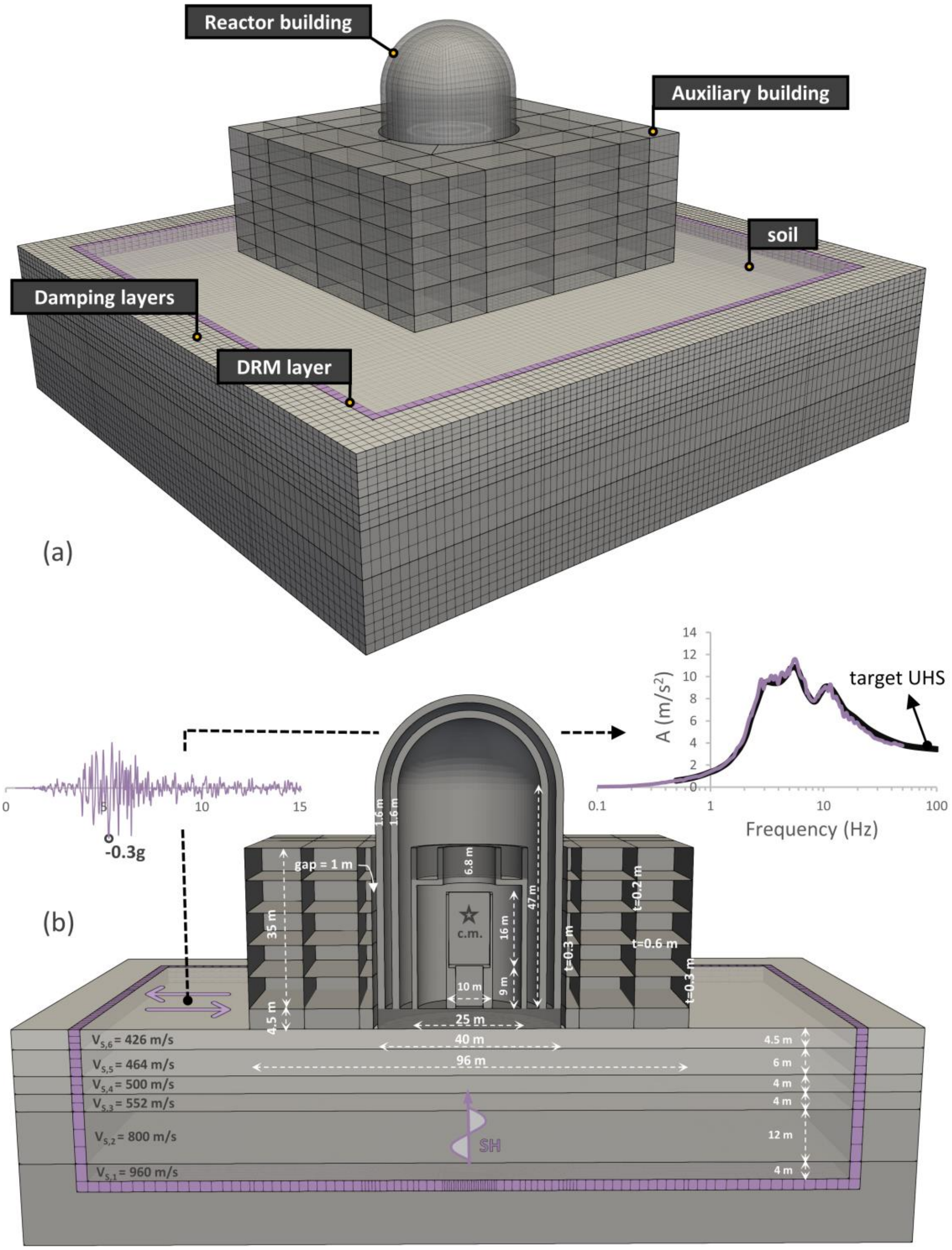

**Figure 1.** (a) Overview of the 3D DRM FE model of the NPP; and (b) cross-section of the model, showing key geometric and material properties. An artificial accelerogram is targeted at the ground surface using the DRM.

## 2.2 NPP Reactor building

The 71.5 m tall NPP reactor building is located at the center of the model, with its containment structure comprising two reinforced concrete (RC) shells; one outer shell and a pre-stressed inner shell. Each shell is 1.6 m thick, with outside radii of 20 m and 16.8 m for the external and internal shells, respectively. A Cylindrical RC Wall (CW) of 35 m height and 12.5 m radius is inside the reactor building, featuring a Water Pool (WP) on its top. In addition, in the center of the reactor building there is a 16 m high Biological Shield Wall (BSW), within which the Reactor Vessel (RV, here not explicitly modeled) sits on a 9 m high Reactor Vessel Pedestal (RVP). Both structures are simplistically assumed to be a square prism made of RC. The BSW and the RVP are 0.5 and 2 m thick, respectively. For simplicity, the two critical components (BWS and RVP) of the reactor building structure associated with the dynamic behavior of the reactor vessel are simply referred to as the Reactor Vessel (RV). The reactor building is supported by a circular 4.5 m thick RC foundation.

The reactor building is modeled using linear elastic 8-noded brick elements with typical RC material properties: Young's modulus $E_{RC}$ = 35 GPa; Poisson's ratio $\nu_{RC}$ = 0.17; and unit weight $\rho_{RC}$ = 2.55 Mg/m$^3$. To roughly account for the additional mass of the heavy permanent equipment on the foundation level and of the water inside the water pool, the unit weight of the foundation and the water pool is increased to 5 Mg/m$^3$, as indicated by the darker grey color in the corresponding part of the reactor building in **Figure 1b**. As a result, the center of mass of the reactor building is at 25 m above the ground surface, while its total weight is approximately 100'000 Mg. A typical 7% viscous Rayleigh damping is assigned to the outer containment wall, and 4% to the inner wall, which is assumed to be prestressed—hence the lower damping value [37].

As a first step in understanding the dynamic behavior of the NPP reactor building, an eigenvalue analysis of the fixed–base reactor building is performed. **Figure 2** displays the fundamental translational vibration modes of various components of the reactor building, along with their corresponding eigenfrequencies. The eigenfrequencies of the first translational vibration modes of the containment walls, the CW, and the RV, which are of significant interest in this study, are approximately 4, 5, and 12 Hz, respectively. Additionally, the second translational vibration modes of the containment walls and the RV are also presented for reference. To assess the adequacy of the current FE mesh discretization, which employs linear 8-noded brick elements, the same reactor building model was discretized with the more accurate—but computationally more expensive—quadratic 27-noded brick elements. The deviation from the translational vibration modes of interest displayed in **Figure 2** was found to be less than 5%, indicating that the linear 8-noded brick elements provide sufficient accuracy.

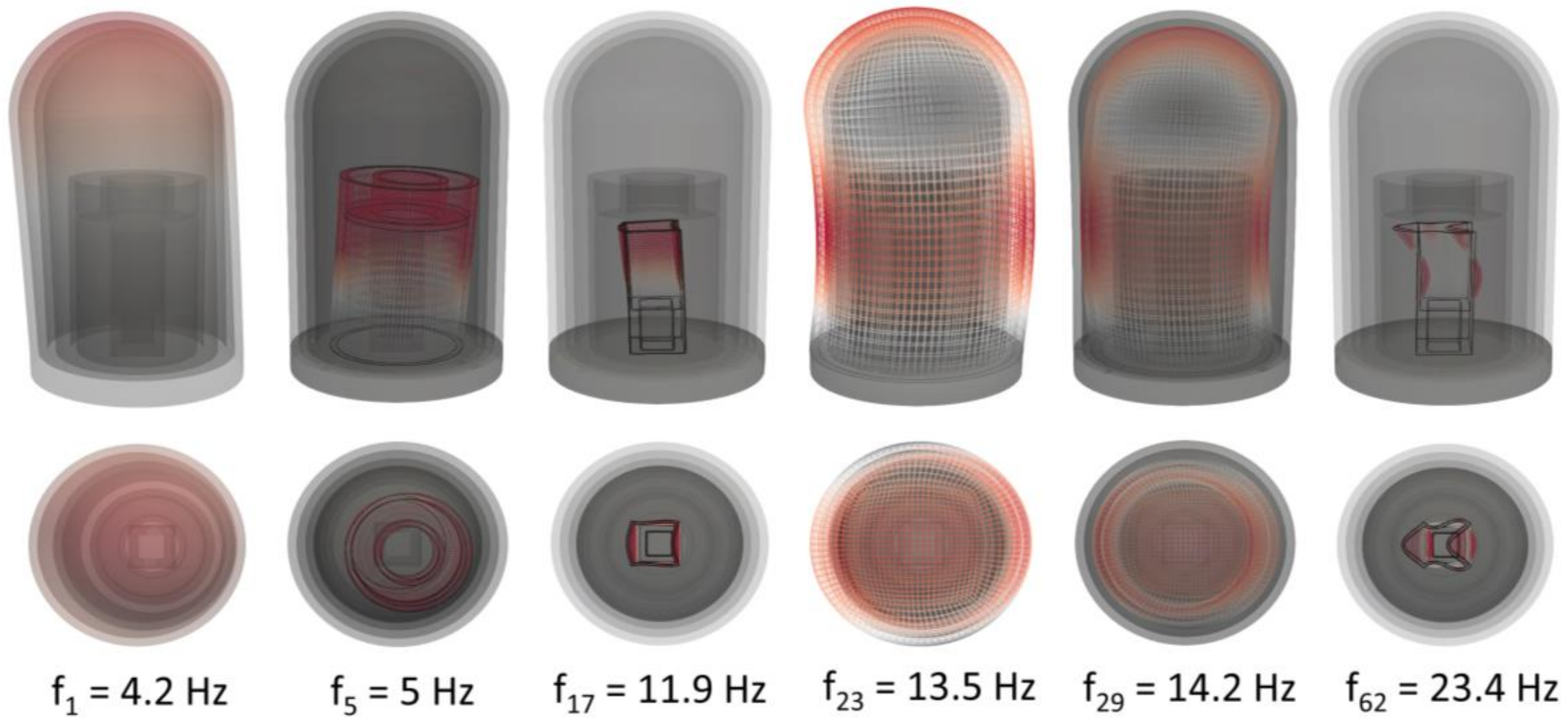

**Figure 2.** Side views (top) and top views (bottom) of the fundamental translational eigenfrequencies of the fixed-base case of the reactor building, and the corresponding deformed shapes.

## 2.3 Auxiliary Building

Typically located adjacent to the reactor building, auxiliary buildings are essential structures for the safe and efficient operation of NPPs, as they house important equipment, such as radioactive waste systems, chemical and volume control systems, and emergency cooling water systems (https://www.nrc.gov/reading-rm/basic-ref/glossary/auxiliary-building.html). In this particular study, a 4-storey, 39.5 m tall and 96 m wide square RC auxiliary building surrounds the reactor containment structure (**Figure 1**), founded on a 4.5 m thick RC foundation. The circular opening in the center is of 21 m radius to fit the reactor building. The internal walls, slabs, and external walls are 0.2 m, 0.6 m, and 0.3 m thick, respectively.

Except for its foundation, the auxiliary building structure is modeled using linear elastic 4-noded shell elements with 6 degrees-of-freedom (dofs) per node. The foundation consists of linear elastic 8-noded brick elements with 3 dofs per node. To ensure that the walls are fixed at the foundation, a master-slave kinematic constraint is applied on the translational dofs of the nodes between the foundation and the embedded walls. All elements of the auxiliary building are assigned the same RC material properties as the reactor building (except for the embedded portions of the walls, which have almost zero unit weight). The auxiliary building has a total weight of about 180'000 Mg and is assigned a Rayleigh damping of 7%.

**Figure 3** presents the first three vibration modes of the fixed-base auxiliary building. The focus of this study is on the first translational mode, which has an eigenfrequency of about 6 Hz. The second torsional vibration mode and the third one, which involves slab vibration, are less relevant as the interaction between the two buildings is mainly through their translational and rocking vibration modes, as demonstrated later on.

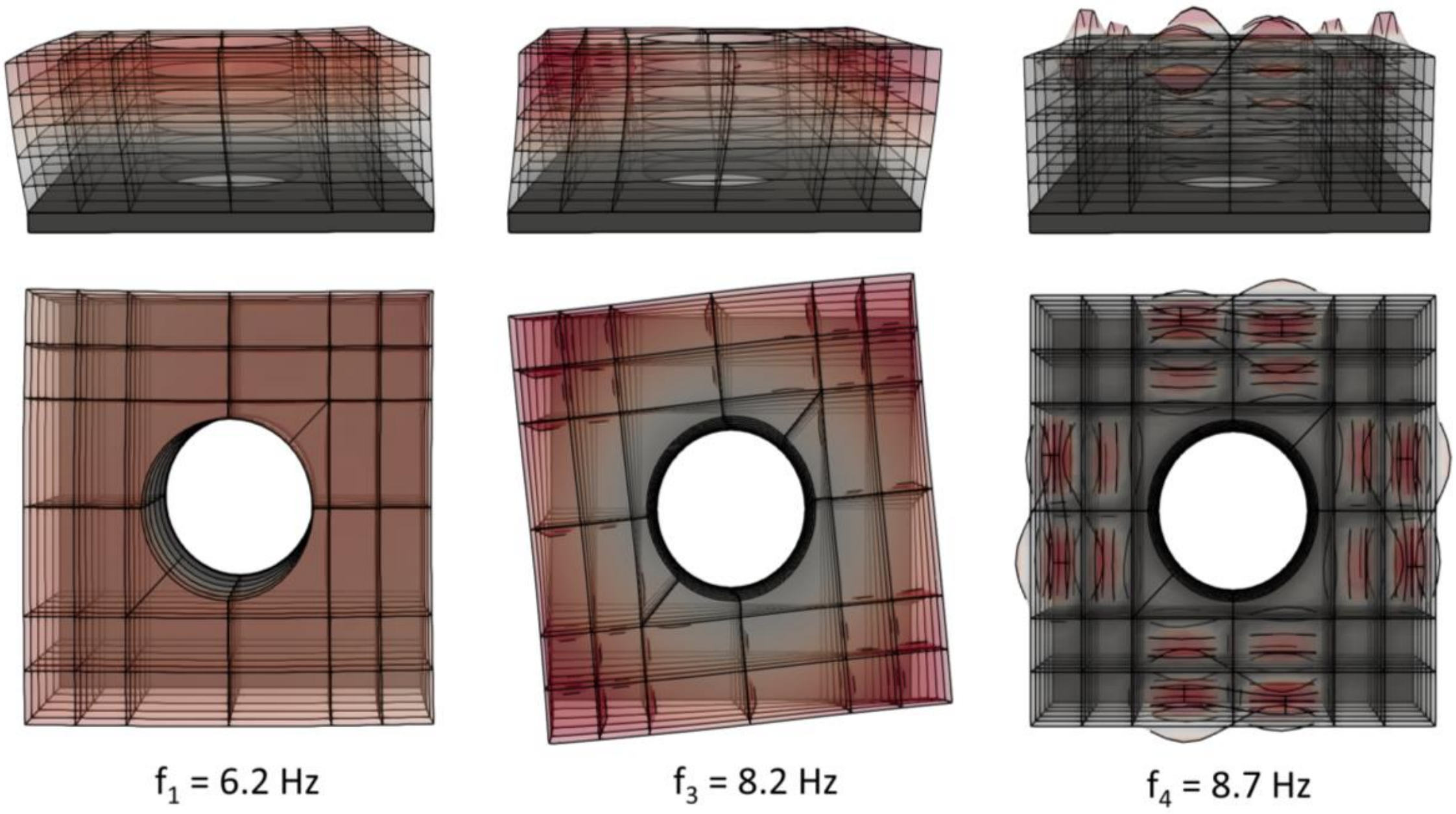

**Figure 3.** Side views (top) and top views (bottom) of the first three eigenfrequencies of the fixed-base case of the auxiliary building, and the corresponding deformed shapes.

## 2.4 Soil

### *Linear elastic soil*

A realistic soil profile, consisting of six layers of increasing shear wave velocity with depth, representative of a Central-European NPP site [38] is selected. The shear wave velocity $V_S$, the unit weight $\rho$, and the Poisson's ratio $\nu$ of each soil layer are summarized in **Table 1**. Linear elastic 8-noded brick elements are employed to model the soil, with at least 10 elements per wavelength [39], assuming that the maximum frequency of interest is 20 Hz. A Rayleigh damping of 2% is assumed, which is typical for linear elastic soil conditions [40].

**Table 1.** Elastic material properties of the layered soil profile.

| Soil Layers | $V_S$ (m/s$^2$) | $\rho$ (Mg/m$^3$) | $\nu$ (-) |
|---|---|---|---|
| 6 | 426 | 2 | 0.4 |
| 5 | 464 | 2 | 0.4 |
| 4 | 500 | 2 | 0.4 |
| 3 | 552 | 2.2 | 0.4 |
| 2 | 800 | 2.2 | 0.4 |
| 1 | 960 | 2.65 | 0.4 |

Therefore, the selected mesh discretization is theoretically able to simulate the propagation of vertical shear waves of up to 20 Hz. To verify this claim, a simplified analysis of a soil column with the same soil layers and mesh discretization as the main model is conducted. Indeed, employing the DRM, the targeted acceleration time history and pseudo-acceleration spectrum (SA) (denoted as input in **Figure 4**) can be successfully reproduced at the ground surface (denoted as output in **Figure 4**) even for up to 30 Hz.

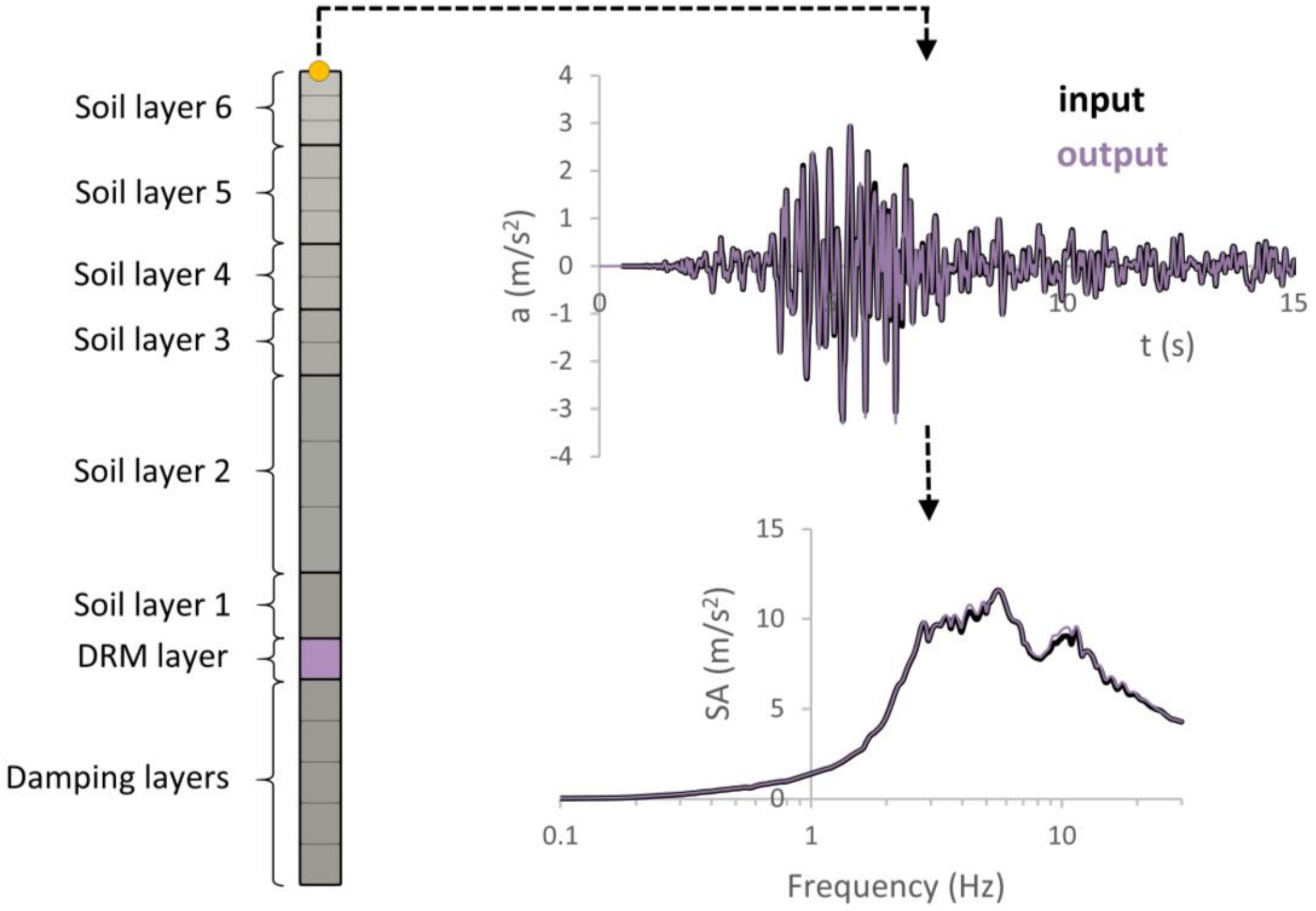

**Figure 4.** Linear elastic soil column: Verification of vertical wave propagation (1-component SH waves) using the DRM. Comparison of the requested (input) to the resulting (output) acceleration time histories at the ground surface (top) with their corresponding spectral acceleration plots (bottom).

## *Nonlinear soil*

Linear elastic or equivalent linear approaches are considered a good starting point for understanding complex dynamic SSI problems. Such methods have been widely used in both research and practice over the past decades, mainly thanks to their ease of use and their reduced computational cost. However, it is well known that soil behavior is inherently nonlinear [41], even at very small strains. Furthermore, as will be shown later on, ignoring nonlinear response does not always lead to conservative results. Therefore, it is crucial to investigate the dynamic response of such important structures using a nonlinear soil model. [42]

In this context, an elastoplastic constitutive model is employed, incorporating a von Mises failure criterion, the Armstrong–Frederick nonlinear kinematic hardening [43], and an associated flow rule. Such model is relatively simple but robust and effective in simulating pressure-independent materials, such as clays. Similar constitutive models have been broadly adopted in research in recent years [44,45]. **Table 2** summarizes the parameters of the constitutive model for all soil layers, where $\rho$, $E$, and $v$ are following **Table 1**; $R$ is the radius of the deviatoric section of the von Mises yield surface that controls the size of the elastic region; and $h_a$ and $c_r$ are two parameters that control the post-yield hardening behavior, with their ratio $h_a/c_r$ being calibrated to achieve the target undrained shear strength $S_u$, which is not a parameter of the model. The target $S_u$ for each layer is defined assuming that the ratio of shear modulus to undrained shear strength $G_0/S_u$ is equal to 800—a typical value [46]—except for the surface layer, which is usually improved, and hence a lower value of $G_0/S_u$ = 500 is chosen. In addition to the hysteretic damping stemming from nonlinear soil response (plastic energy dissipation), a viscous Rayleigh damping of 2% is assumed to crudely account for the viscous energy dissipation caused by the soil–pore fluid interaction.

**Table 2.** Input parameters of the Von Mises elastoplastic model with Armstrong-Frederick nonlinear kinematic hardening for each soil layer, along with the corresponding undrained shear strength $S_u$.

| Soil Layers | $E$ (MPa) | $\rho$ (Mg/m$^3$) | $\nu$ (-) | $R$ (kPa) | $h_a$ (Mpa) | $c_r$ (-) | $S_u$ (kPa) |
|---|---|---|---|---|---|---|---|
| 6 | 1016.3 | 2 | 0.4 | 8 | 2000 | 1600 | 725 |
| 5 | 1205.7 | 2 | 0.4 | 7 | 2000 | 2162 | 538 |
| 4 | 1400 | 2 | 0.4 | 8 | 2300 | 2150 | 625 |
| 3 | 1877 | 2.2 | 0.4 | 11 | 3100 | 2150 | 838 |
| 2 | 3942.4 | 2.2 | 0.4 | 23 | 6500 | 2150 | 1760 |
| 1 | 6838.3 | 2.65 | 0.4 | 40 | 11000 | 2150 | 3053 |

To validate the employed constitutive model at the element level, a single element of each soil layer is subjected to cyclic shear loading of increasing amplitude. For each layer, the computed normalized secant shear modulus – shear strain ($G/G_0 - \gamma$) curve is compared with experimental data from the specific site [38]. Such a comparison is illustrated in **Figure 5**, which compares the numerical and experimental $G/G_0 - \gamma$ curves for the 2$^{nd}$ soil layer. Despite its fluctuations, the numerical curve remains within the upper and lower bounds of the experimental curve for the specific site, for the entire strain range. Almost identical plots were obtained for all soil layers, validating the selected parameters of the constitutive model. Finally, to estimate the expected level of inelasticity in each soil layer due to the propagating shear waves, a nonlinear soil column is modeled and subjected to the previously discussed artificial acceleration time history. The resulting vertically propagating shear waves (SH) entering the model reproduce the target acceleration time history at the top of the soil. **Figure 6** depicts the shear stress – shear strain ($\tau - \gamma$) loops for each soil layer. The maximum shear strain observed is slightly less than 3x10$^{-4}$, which can be considered a fairly inelastic behavior, corresponding to $G/G_0$ of about 0.6. However, it should be noted that a higher level of inelasticity is expected to take place in the soil under the edges of the reactor and auxiliary building structures during their oscillation.

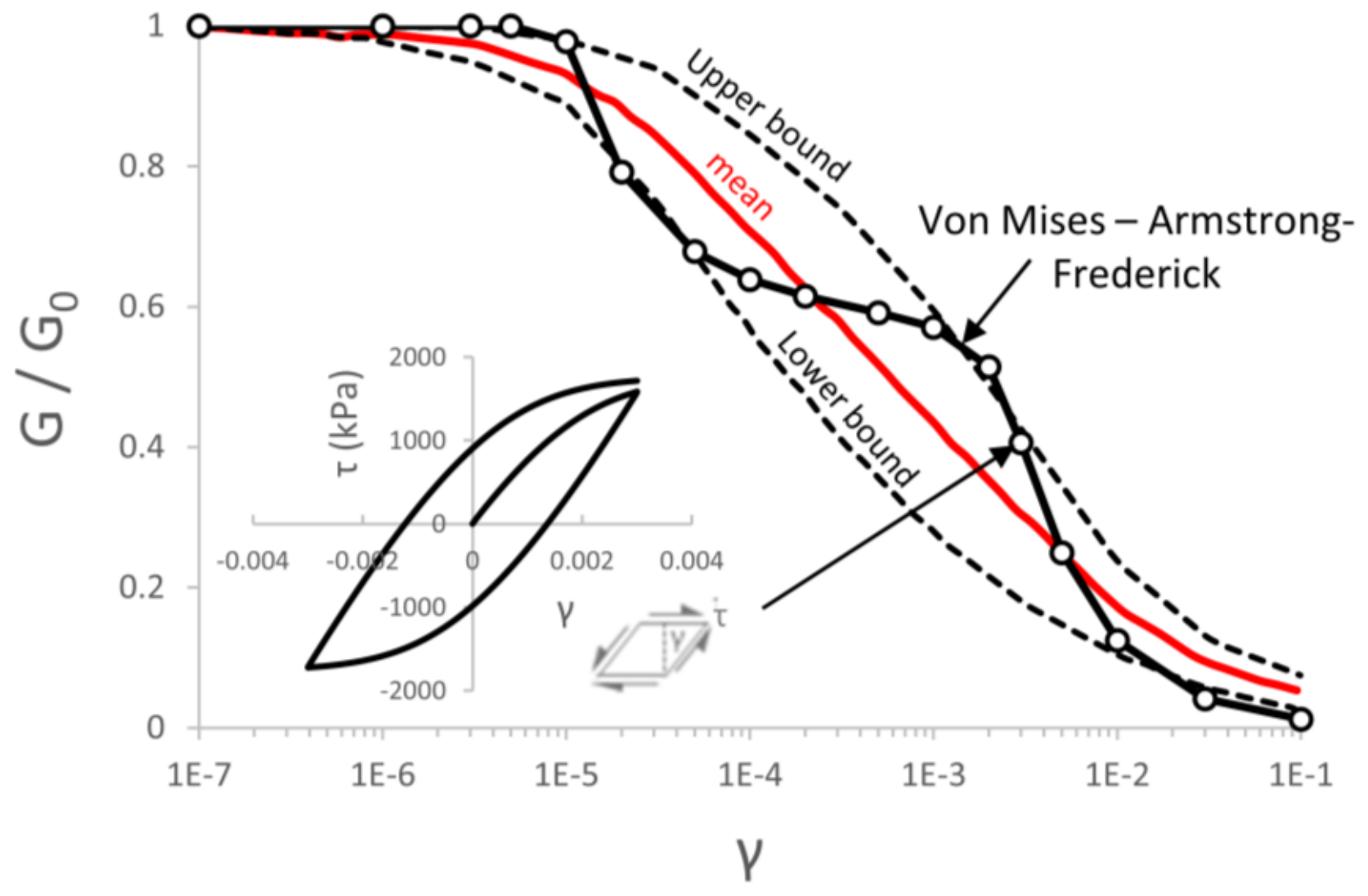

**Figure 5.** Nonlinear soil response modelled with a von Mises–Armstrong-Frederick soil model, for $G_0/S_u$ = 800. Comparison of computed $G/G_0 - \gamma$ degradation curves with the experimental curves from the specific site for soil layer 2. The inset plots the shear stress–strain ($\tau - \gamma$) loops for $\gamma$ = 0.003. Nearly identical plots are obtained for the remaining soil layers.

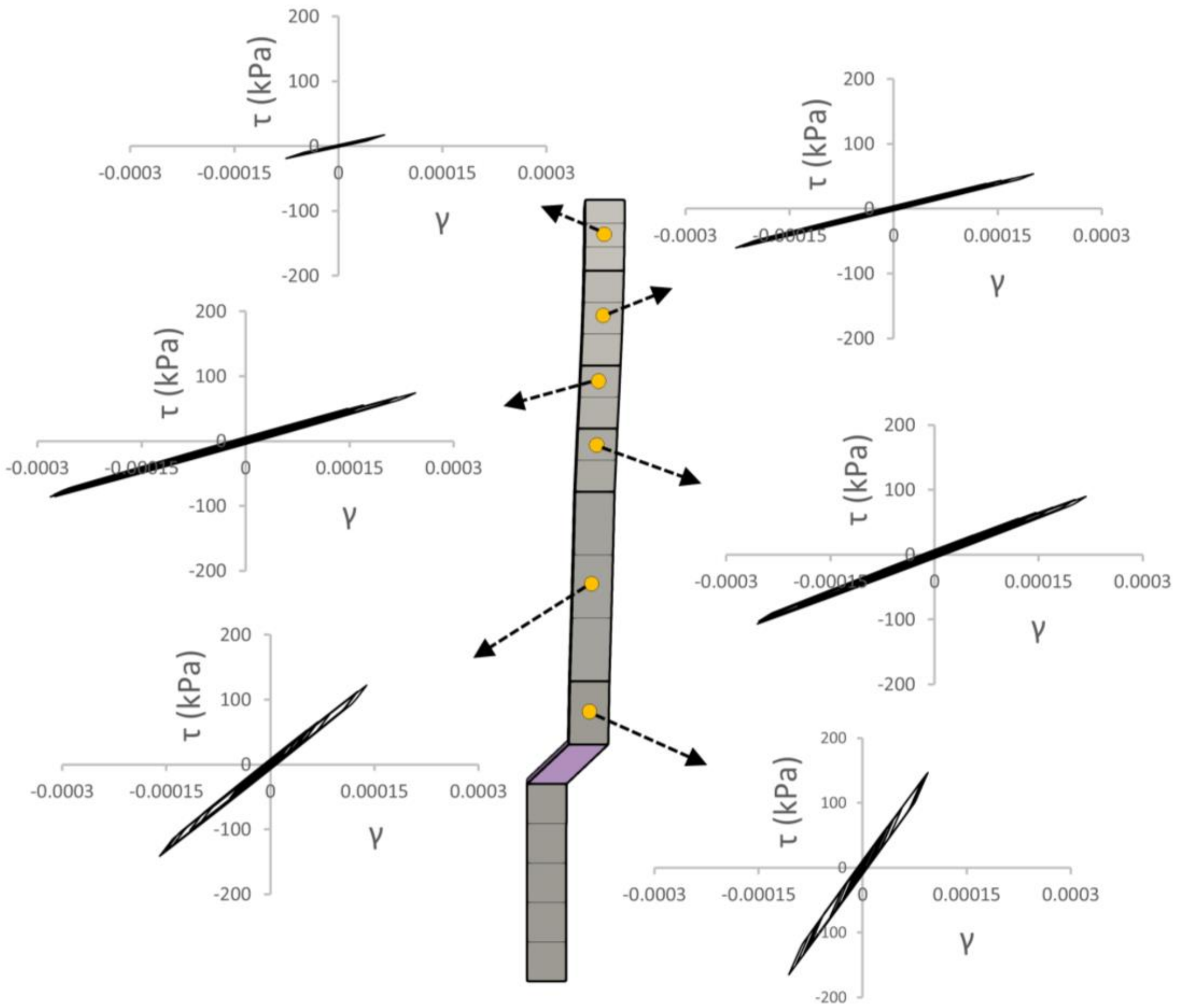

**Figure 6.** Dynamic analysis of nonlinear soil column: shear stress–strain ($\tau - \gamma$) loops for each soil layer.

## 2.5 Interfaces

In the simplest elastic model, all interfaces (reactor building–soil and auxiliary building–soil) are modeled as tied. However, during strong seismic shaking, the soil–structure interfaces may be subjected to detachment and sliding, which can alter the dynamic response of the structures. As it will be demonstrated later, this may not necessarily result in a reduction of seismic response compared to the tied interfaces. Real-ESSI provides several types of interface elements that can capture different aspects of soil–structure interaction with increasing complexity. A zero-length, stress-based, dry contact element with nonlinear hardening shear behavior is adopted for the present study, called StressBasedSoftContact_NonLinHardShear [47,48]. Such contact model allows simulation of interface behavior in both normal and tangential directions to the contact surface. In the normal direction, a nonlinear elastic penalty stiffness function representing a soft contact with stiffness increasing exponentially with penetration is used to model the contact behavior (***Figure 7*a**). This is considered a realistic representation of the normal soil–structure interface behavior, as the normal contact force changes gradually upon contact and becomes zero upon detachment. For the nonlinear contact behavior in the tangential direction, an Armstrong-Frederick nonlinear hardening model is used, where the shear stress to normal stress ratio ($\mu = \tau/\sigma_n$) increases nonlinearly from 0 to the value of the residual friction coefficient $\mu_r$ (***Figure 7*b**).

Due to the lack of experimental data for the NPP soil–structure interfaces, reasonable engineering assumptions were made to calibrate the parameters of the interface model. The input parameters used in the model are listed in **Table 3**, where $k_{n0}$, $sr$, and $k_{n,max}$ correspond to the initial normal stiffness, the normal stiffness stiffening rate, and the maximum normal stiffness, respectively. These parameters control the stiffness behavior of the interface in the normal direction and result in negligible penetration, of the order of 1x10$^{-6}$ m. The shear stiffness behavior of the interface is controlled by $k_t$, the shear stiffness at 101 kPa normal stress, and $\mu_r$, the residual friction coefficient, which is set to 0.6. A simple example of shear behavior between two elastic brick elements subjected to static cyclic shear excitation, after application of a vertical stress of 600 kPa (which is representative of the vertical stress below the foundation of the reactor building) is shown in **Figure 7b**. It is essential to use a sufficiently high value of $k_t$ to reach the residual friction coefficient after reasonable shear displacement $\delta_t$ (here approximately 0.3 mm). However, excessively high values of $k_t$ could lead to numerical convergence problems, so a compromise is required. To support numerical stability and to serve as a physical energy-damping mechanism, the interface model provides two additional parameters $c_n$ and $c_t$, which correspond to the viscous damping in the normal and tangential directions, respectively. These parameters represent the viscous energy dissipation that can occur at the soil–structure interface during the opening and closing of the gap, as the fluids (air or water) interact with the structure. Parametric analyses indicated that a shear viscous damping of 4 kPa*s and a normal viscous damping of 0.5 kPa*s provide numerical convergence without compromising the reliability of the results for the studied problem. The final physical parameter of the interface model is the shear zone thickness $h$, which is set to 1 mm, being typically assumed to be 5–10 times the mean soil particle diameter $D_{50}$ [49].

**Table 3.** Input parameters of the nonlinear interface model.

| $k_{n0}$ (MPa) | $sr$ (-) | $k_{n,max}$(MPa) | $k_t$ (kPa) | $c_n$ (kPa*s) | $c_t$ (kPa*s) | $h$ (mm) | $\mu_r$ (-) |
|---|---|---|---|---|---|---|---|
| 500 | 100 | 2500 | 1000 | 0.5 | 4 | 1 | 0.6 |

In the absence of experimental data to properly calibrate the interface model, the chosen parameters were derived using a combination of engineering expertise and parametric analyses, assessing the impact of each parameter on the structural response for the studied problem. As such, these parameters should not be blindly applied in future studies. To obtain a comprehensive understanding of the interface model described in this paper, the readers are encouraged to refer to Sinha [48].

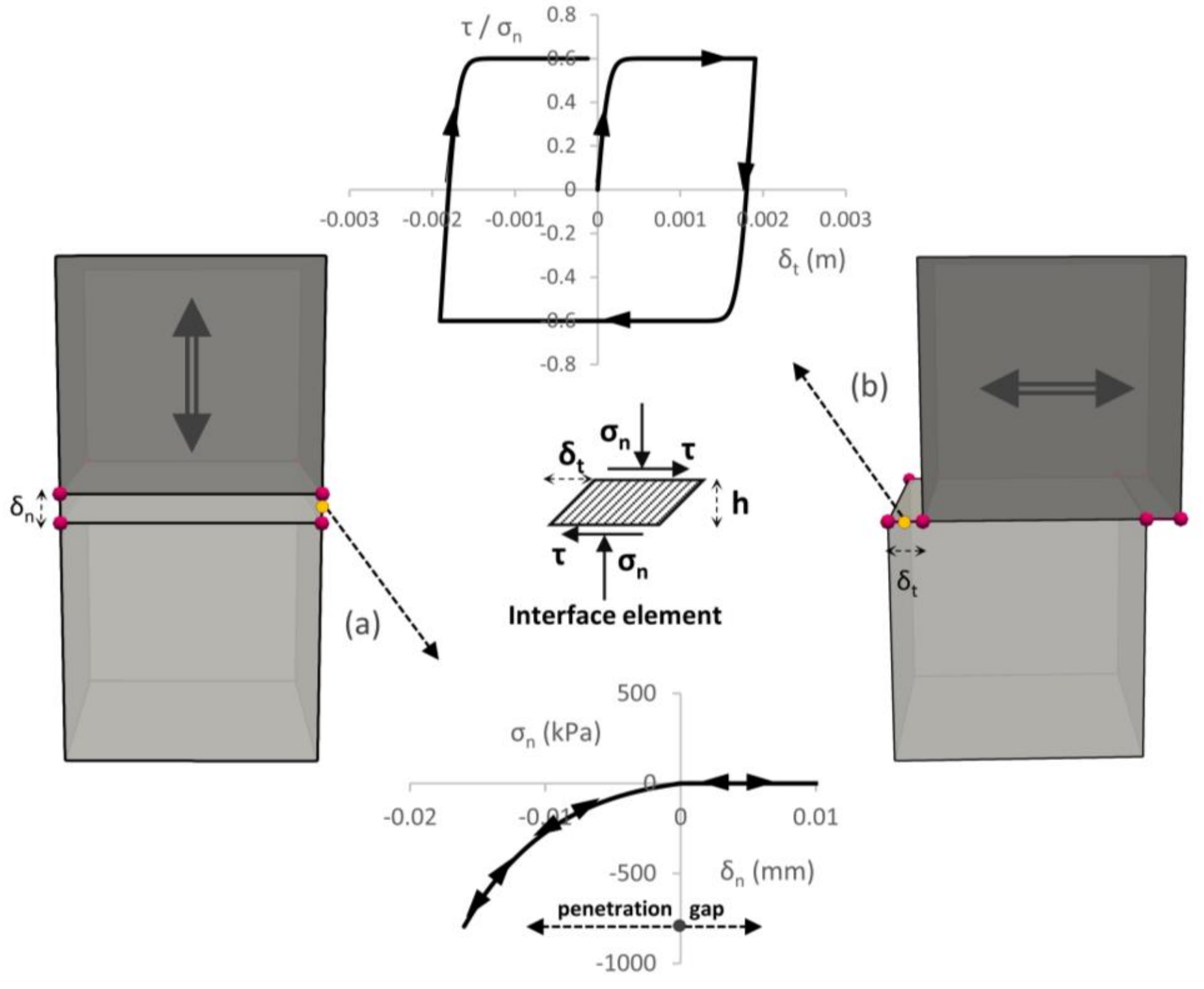

**Figure 7.** Cyclic response of the nonlinear interface elements used in this study: (a) normal stress–normal displacement ($\sigma_n - \delta_n$); and (b) shear stress over normal stress ratio–shear displacement ($\tau/\ \sigma_n - \delta_t$) response.

# 3 Elastic soil and Tied interfaces

## 3.1 The effect of SSI

SSI effects are often neglected during the design or assessment of structures, assuming that it is beneficial. This assumption is based on the expectation that the soil will absorb a significant portion of the seismic energy, primarily through its material and radiation damping, that would have otherwise been absorbed by the structure. However, a structure sitting on a flexible soil base will experience a modification of its vibration modes, and depending on the structure and the seismic excitation, these new vibration modes could be detrimental. To illustrate this, **Figure 8** compares the response of a fixed-base model of the NPP reactor building, in which the selected motion is applied at its base, with the full SSI model where the DRM is used to target the same motion at the ground surface. The top node acceleration time history responses of the external containment walls are plotted at the top (**Figure 8b**), accompanied by their respective elastic response spectra at 5% damping (SA) at the bottom (**Figure 8c**). The peak of the black curve (no SSI) at around 4 Hz corresponds to the first vibration mode of the external walls of the eigenvalue analysis (see **Figure 2**). Accounting for elastic SSI leads to a dramatic reduction of this peak—hence the general belief that neglecting SSI is conservative—but at the same time, a new vibration mode is introduced at around 2 Hz, which is related to the rocking motion of the NPP reactor building as a tall rigid body on a soft base (**Figure 8a**). Such rocking motion cannot be predicted by assuming a fixed-base model. However, rocking of the reactor building could potentially compromise its safety, calling for a proper investigation that accounts for SSI.

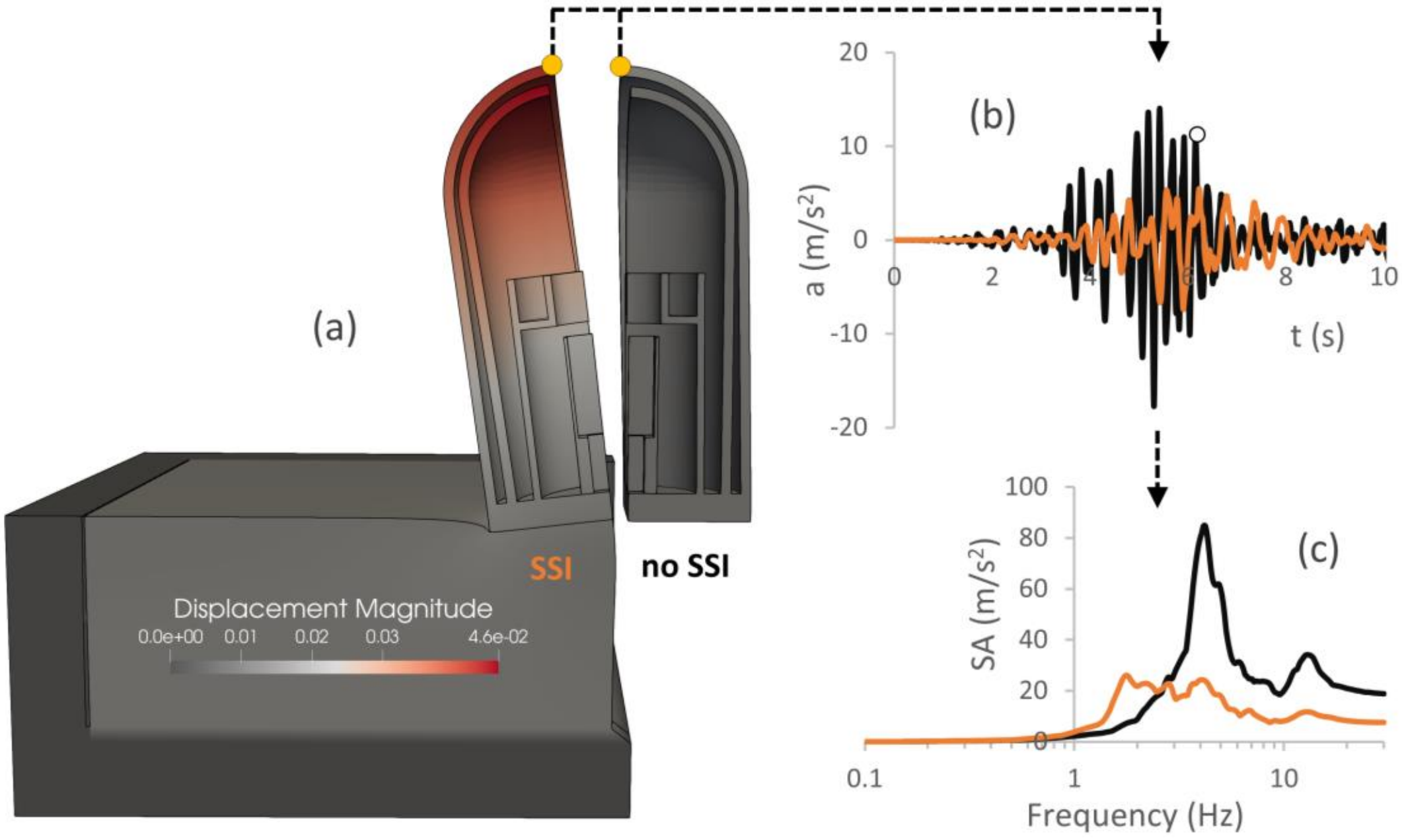

**Figure 8.** The effect of elastic SSI on the response of the external wall of the reactor building. Comparison of the fixed-base model to the one that accounts for elastic SSI: (a) snapshot of displacement contours at *t* = 6.2 s (scale factor 300); (b) acceleration time histories of the top node of the external walls; and (c) the corresponding elastic acceleration response spectra.

## 3.2 The effect of the auxiliary building on the reactor building (SSSI)

If SSI effects are often neglected, SSSI effects are almost never considered in practice. As mentioned earlier, the main goal of this study is to explore the significance of SSSI modeling, which can have either positive or negative effects on the response of the structure. To this end, the dynamic effect of the auxiliary building on the reactor building is first studied, assuming tied interfaces and elastic soil conditions. **Figure 9** presents the elastic response spectra at 5% damping (SA) for seven critical points of the reactor building, with and without the presence of the auxiliary building (i.e., with and without SSSI). Point 1 is at the center of the foundation, points 2 and 3 are on the RV, point 4 is at the top of the WP, point 5 is at the inner side of the internal containment wall at a height where usually a crane is attached, and points 6 and 7 are at the top of the internal and external containment walls, respectively. The SA at the top of the auxiliary building is also plotted (point 8), having a clear peak close to 4 Hz related to its horizontal translational motion. This is significantly smaller than the 6.2 Hz eigenfrequency of the eigenvalue analysis for the fixed-base case (**Figure 3**), and is due to the flexible soil base. A noteworthy observation in **Figure 9** is the overall increase in SA values caused by the presence of the auxiliary building for the majority of the investigated points of the reactor building, particularly close to the auxiliary's building resonant frequency (see for example point 7 at 4 Hz).

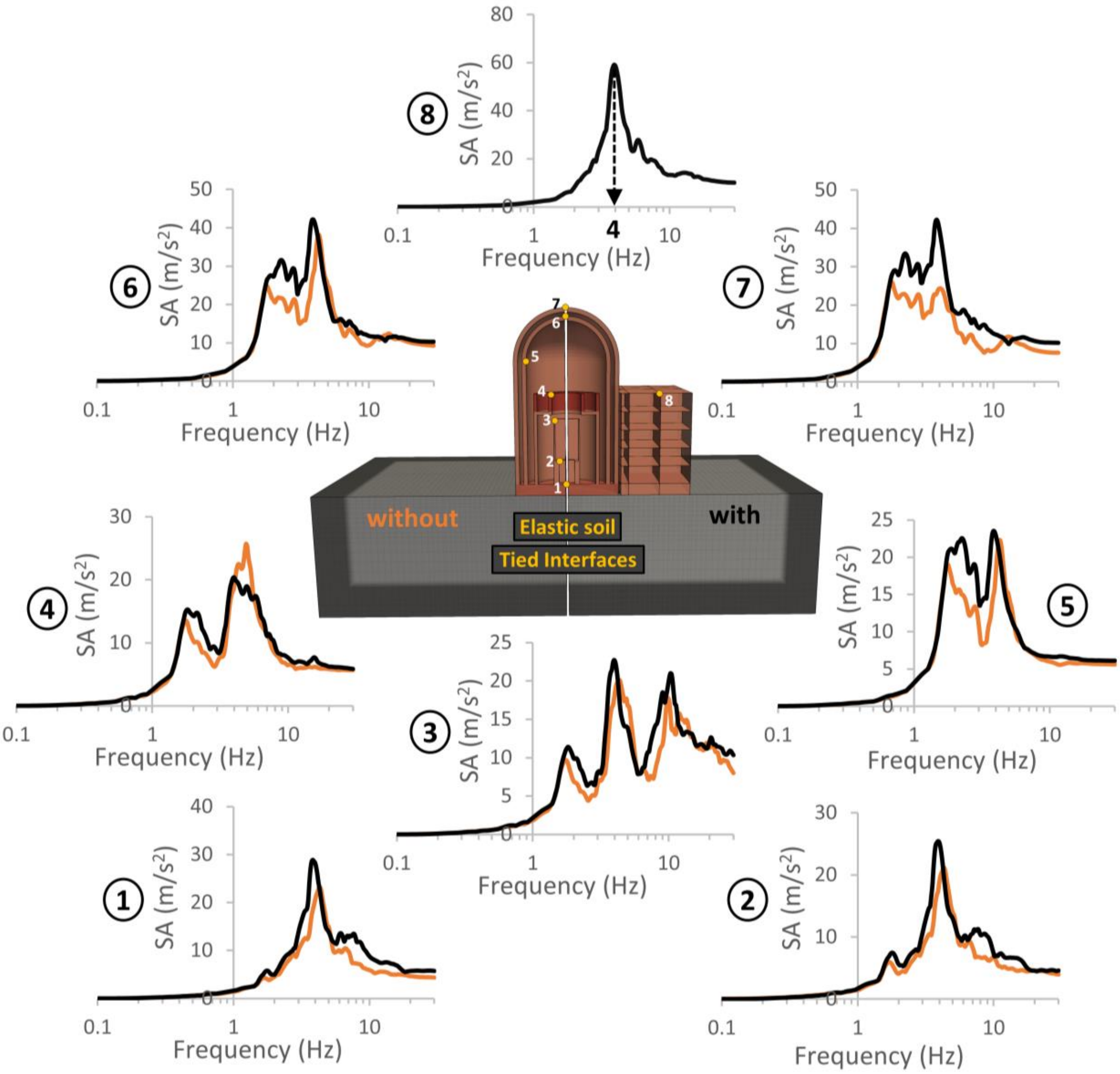

**Figure 9.** Elastic soil and tied interfaces. The effect of SSSI on the response of the reactor building. Comparison of the elastic acceleration response spectra SA at critical points of interest on the reactor building, "with" and "without" the presence of the auxiliary building.

This amplification in the response of the reactor building is attributed to the out-of-phase rocking motion of the two buildings, as clearly shown in **Figure 10** which plots the soil deviatoric effective stress contours for the two cases. Notice the increase beneath the adjacent edges of the foundations of the two structures caused by the additional amount of vertical pressure due to the out-of-phase rocking motion of the auxiliary building. This visibly increases the rotation of the reactor building, leading to acceleration amplifications, particularly for its rocking-prone parts, such as the external containment wall (point 7). This result is consistent with the earlier work of Roesset et al. [11], where it was shown that the rocking of one structure can be significantly altered in the presence of another nearby structure.

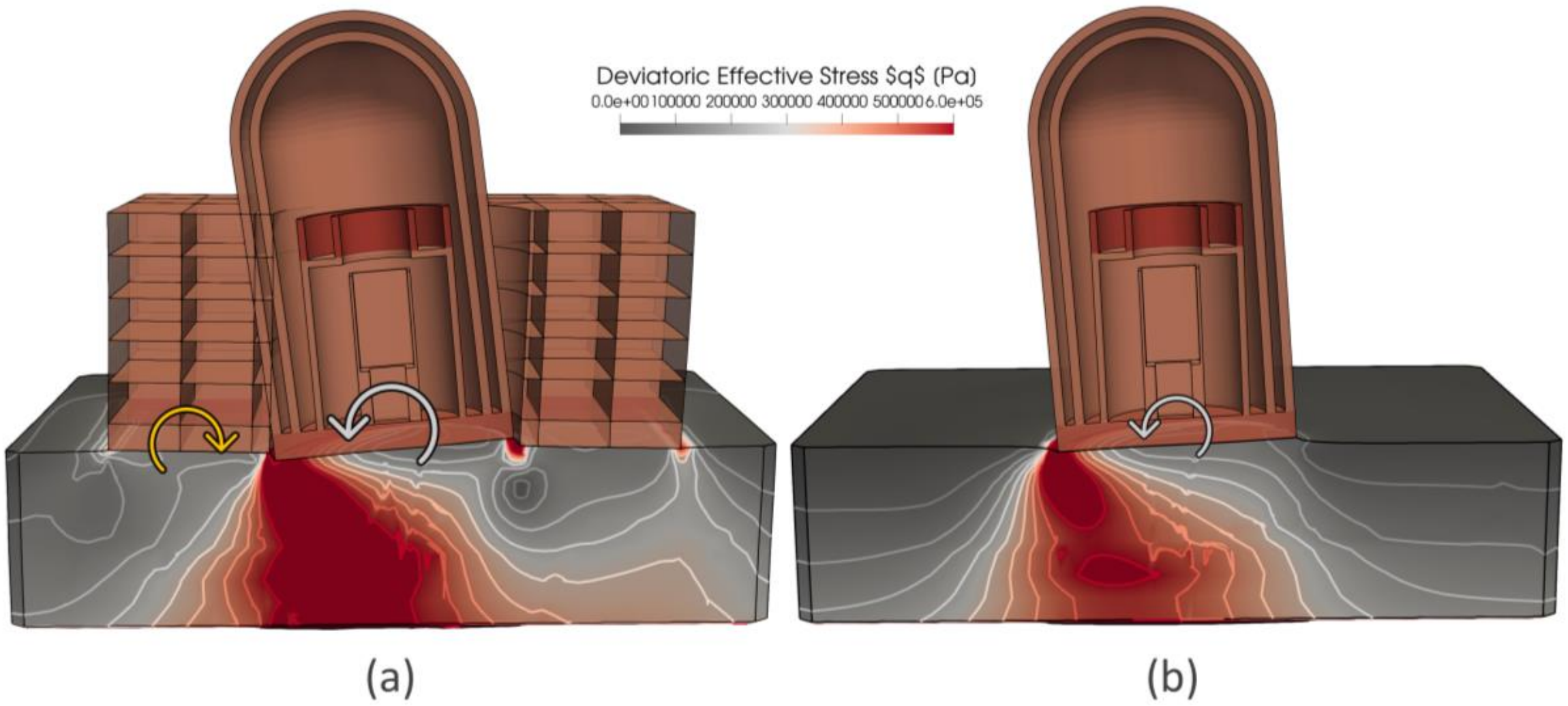

**Figure 10.** Elastic soil and tied interfaces. Snapshots of deviatoric effective stress contours: (a) "with"; and (b) "without" the auxiliary building at *t* = 5.58 s (scale factor 200). The out-of-phase rotational coupling between the two buildings leads to an aggravation of the seismic response of the reactor building.

At this point, it is worth drawing an analogy between the out-of-phase amplification mechanism of the auxiliary building and the out-of-phase response attenuation mechanism of *seismic resonant metamaterials*. As shown by Kanellopoulos et al. [50], the seismic response of a horizontally vibrating building can be reduced by introducing multiple horizontal resonators into the ground adjacent to the building. These resonators are tuned to vibrate out-of-phase with the ground at the frequency of interest (i.e., the resonant frequency of the building). In this case, the out-of-phase horizontal forces from the resonant metamaterials are beneficial to the building, as they oppose the horizontal forces from the seismic waves. Thus, if it were possible here to restrain the rotations and allow only the horizontal translational motion of the auxiliary building and the reactor building, their out-of-phase interaction would actually benefit the reactor building. As it will be shown later, when soil inelasticity is considered, the rocking motion of the soil–buildings system is suppressed once the soil yields, and the horizontal out-of-phase mechanism begins dominating over the rocking one, favoring the beneficial interaction between the two buildings.

# 4 Elastic soil and Nonlinear interfaces

## 4.1 The effect of nonlinear interfaces

The modelling sophistication is enhanced by introducing the previously discussed nonlinear interface at the soil–reactor building and soil–auxiliary building contacts. The effect of the nonlinear interface on the dynamic response of the reactor building is evaluated by comparison with the tied interface cases. Initially, the auxiliary building is omitted to focus on the effect of the nonlinear interface on the response of the reactor building. The nonlinear interfaces lead to a minimal decrease in SA at the 7 characteristic points (not shown here), of 10% or less for frequencies below 10 Hz. This insignificant effect is due to the negligible amount of sliding at the soil–foundation interface (of the order of 1–2 mm), which implies a negligible amount of energy dissipation. Interestingly, the nonlinear interface has a notable effect on the SA of point 3 (top of the RV) at its resonant frequency of 10 Hz (recall from **Figure 2** that its fixed-base resonant frequency is about 12 Hz), causing a significant SA amplification of about 50%. This is depicted in **Figure 11d**, along with cross-sectional views of the compared models (**Figure 11a**) displaying the 10–13 Hz filtered displacement field of interest. Additionally, notice the detachment of the foundation from the soil with its corresponding uplifting time history diagram (**Figure 11b**). As can be visibly inferred by comparing the deformation of the RV for the two cases, this opening and closing of the gap at the soil–foundation interface introduces higher frequency vibrations to the reactor building foundation, leading to the observed amplification of SA at RV's top. The acceleration time history diagrams of the RV's top points (**Figure 11c**) further confirm this observation, showing an increase in the 10–13

Hz filtered acceleration time history plots of the nonlinear interface case after 5.5 s and 6 s, when gap opens and closes twice.

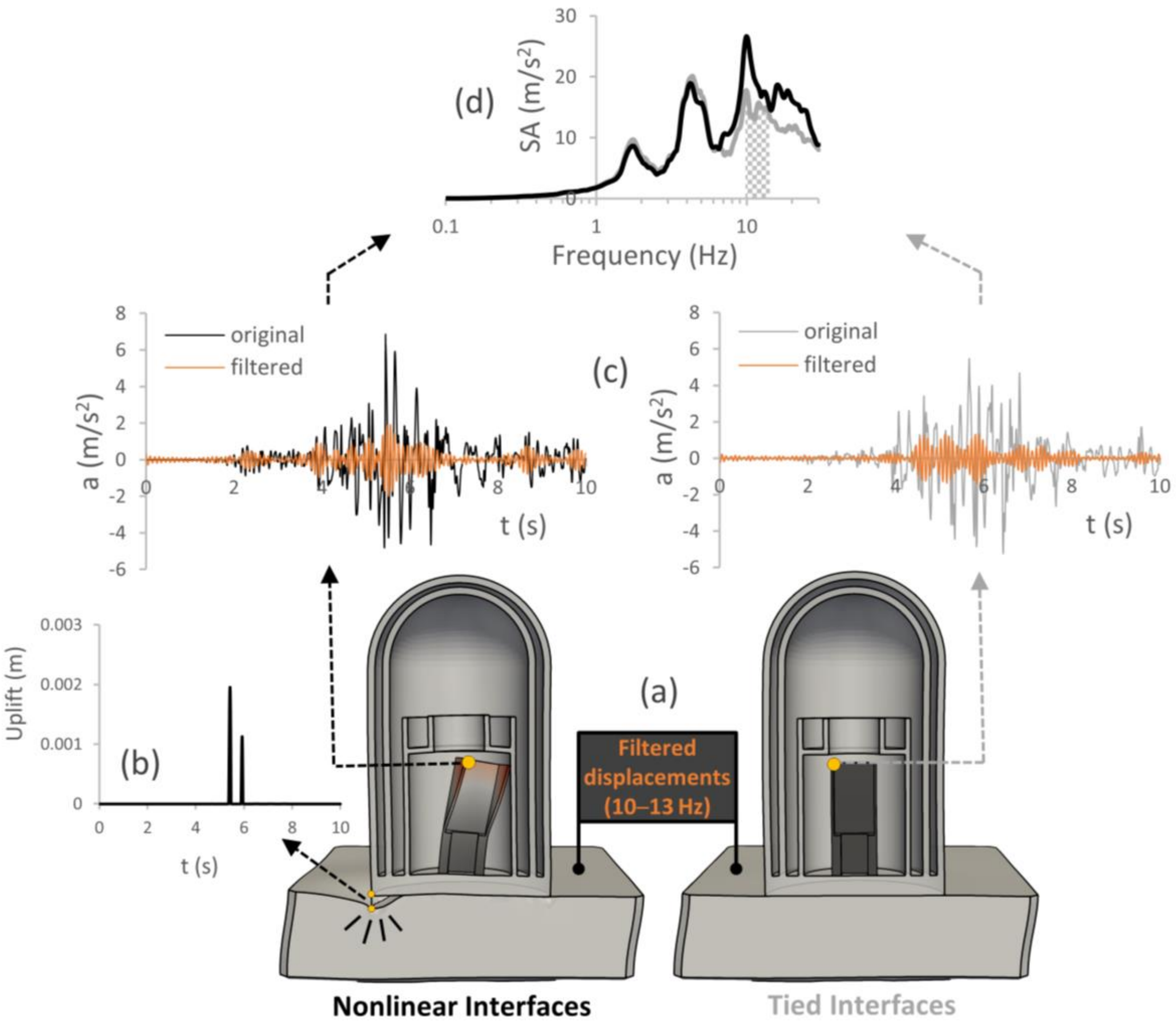

**Figure 11.** The effect of nonlinear interfaces on the RV's response in the absence of the auxiliary building. Comparison with the case of tied interfaces: (a) snapshot of filtered displacement contours (10–13 Hz) at *t* = 5.5 s (scale factor 20'000); (b) time history of uplifting at the left edge of the foundation; (c) filtered and original acceleration time histories of the RV's top; and (d) the corresponding elastic acceleration response spectra.

The next step is to compare the response of the compete 3D model, including the auxiliary building, with a nonlinear and a tied interface. By doing so, the effect of the nonlinear interfaces on the reactor–auxiliary building interaction can be estimated. The cross-sectional views of **Figure 12a** compare the two cases, displaying a decrease of displacement of both buildings when the nonlinear interface is considered. The rotations of the buildings are also visibly decreased due to some sliding. The zoomed-in plots highlight the sliding of the auxiliary building. The reduction of the peak in the SA of point 8, thanks to the energy dissipation at the soil-foundation interface of the auxiliary building, is expected to lead to a reduction of its negative effect on the response of the reactor building (**Figure 12b**). Indeed, the comparison of SA at point 7 (top of the external containment wall), indicates that the nonlinear interface leads to a reduction of the detrimental interaction between the two structures. However, it remains to be seen whether the effect of the auxiliary building on the reactor building is completely reversed (from detrimental to beneficial) or is merely reduced (less detrimental).

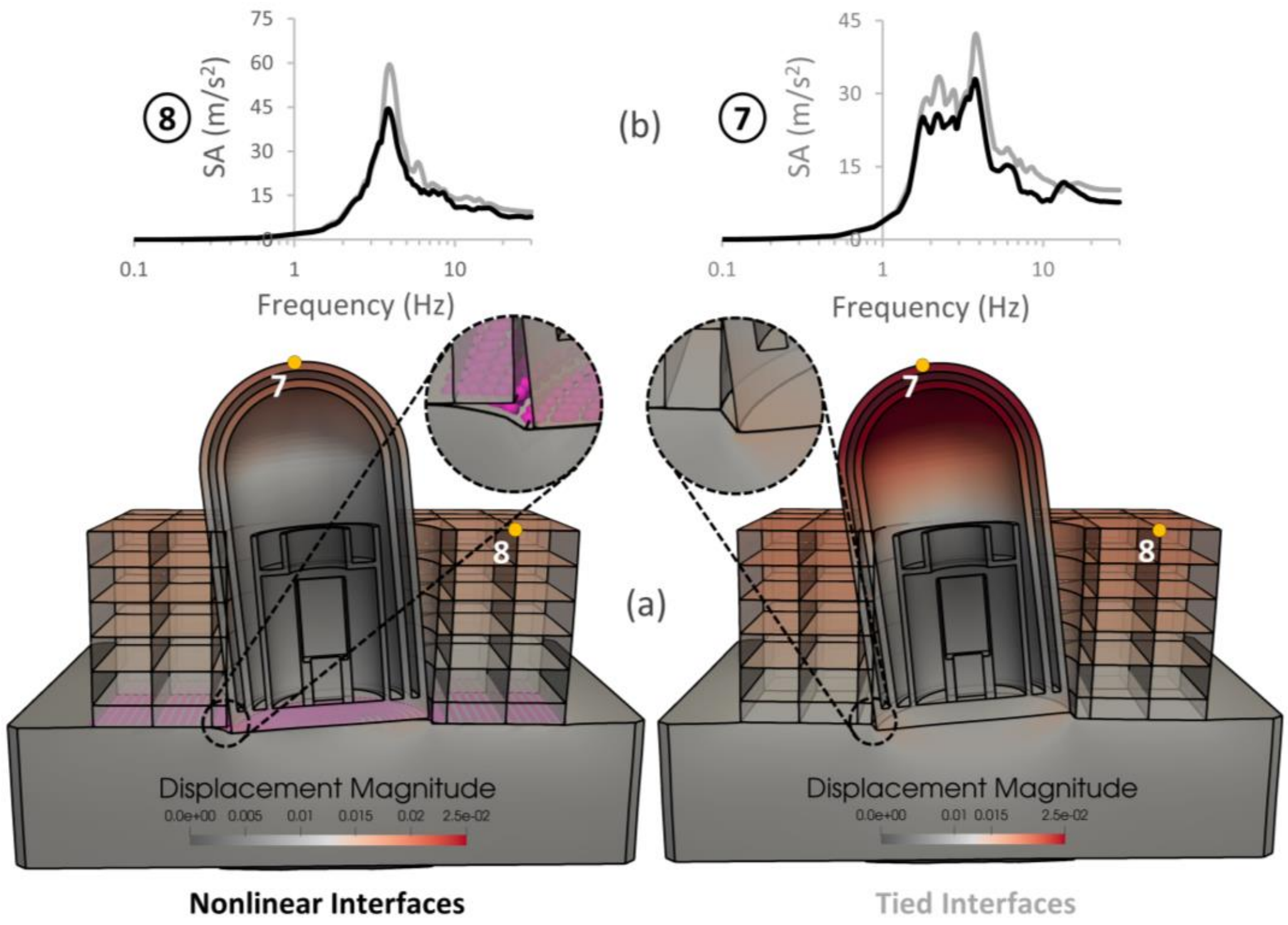

**Figure 12.** The effect of nonlinear interfaces on the RV's response in the presence of the auxiliary building. Comparison with the case of tied interfaces: (a) snapshot of displacement contours at *t* = 5.58 s (scale factor 200); and (b) elastic acceleration response spectra at points 7 and 8. The reduction in SA at point 8 caused by the nonlinear interfaces leads to reduction of the negative effect of the auxiliary building on the response of the reactor building.

## 4.2 The effect of the auxiliary building on the reactor building (SSSI)

The effect of the auxiliary building on the reactor building is assessed by comparing the response of the reactor building in the absence and presence of the auxiliary building, considering nonlinear interfaces for both cases. The SA for the points of interest are compared in **Figure 13**, as previously. As expected, the detrimental effect caused by the auxiliary building is suppressed compared to **Figure 9**, but is still present for frequencies up to 4 Hz. However, the beneficial horizontal out-of-phase interaction, which is enhanced by interface sliding, starts gaining ground over the rocking interaction, as manifested by the noticeable reduction in SA of the reactor building after 4 Hz (see shaded areas). In addition, the presence of the auxiliary building substantially reduces the high-frequency response of the RV's top (point 3). This can be explained with the help of **Figure 14**, which shows the cross-sectional views of the 10–13 Hz filtered displacement fields of the two models, along with the corresponding filtered acceleration time histories of point 3 (top of the RV). Notably, the closing of the gap at the left edge of the soil–foundation interface of the standalone reactor building, which was found earlier to amplify the RV's response at around 10 Hz, is prevented by the opposite motion of the auxiliary building (see zoomed-in plot in **Figure 14a**).

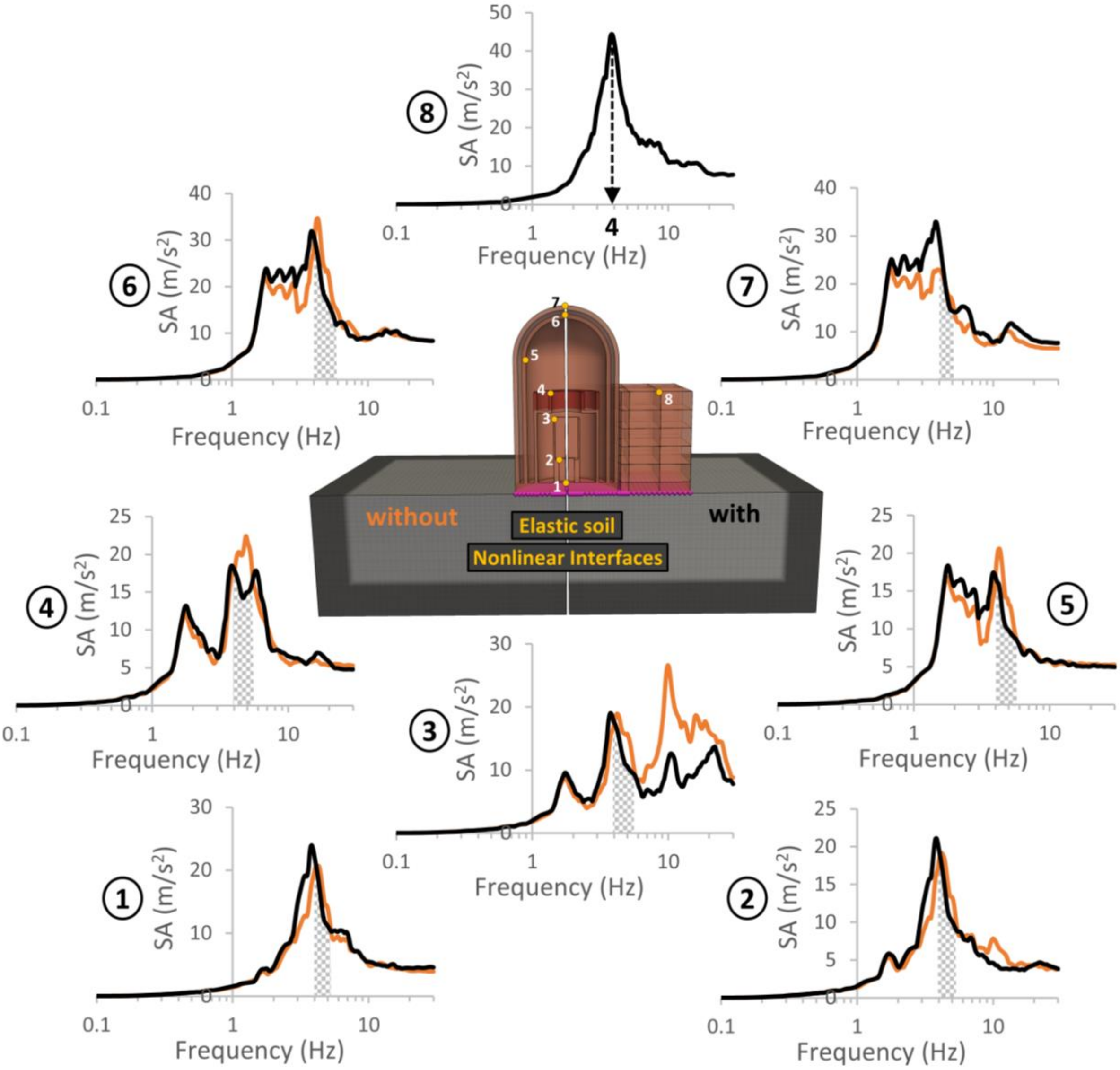

**Figure 13.** Elastic soil and nonlinear interfaces. The effect of SSSI on the response of the reactor building. Comparison of the elastic acceleration response spectra SA at critical points of interest on the reactor building, "with" and "without" the presence of the auxiliary building.

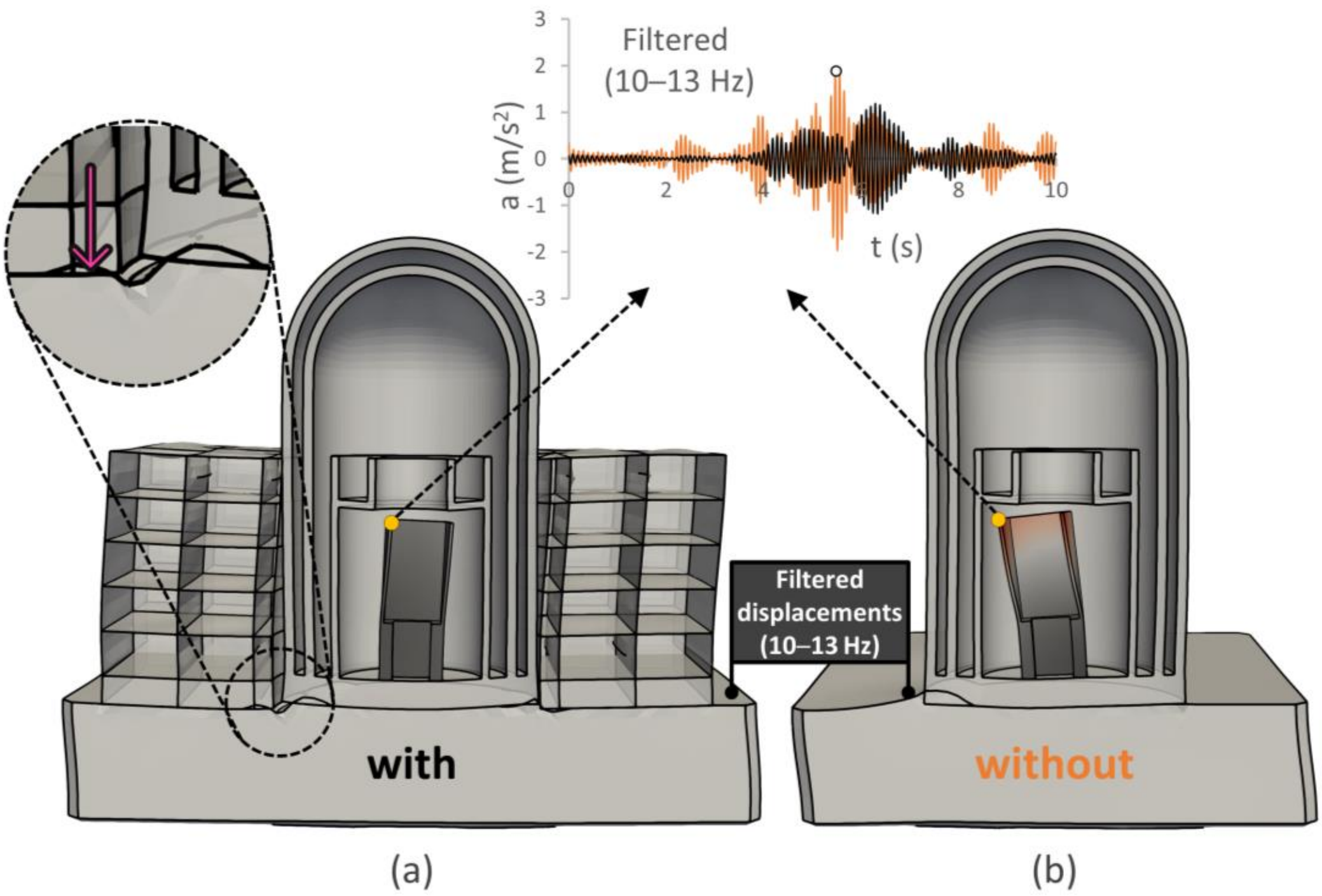

**Figure 14.** Elastic soil and nonlinear interfaces. Snapshot of filtered displacement contours (10–13 Hz): (a) "with"; and (b) "without" the presence of the auxiliary building at *t* = 5.47 s (scale factor 20'000). The inset plots the filtered acceleration time histories at the top of the RV.

# 5 Nonlinear soil and Nonlinear interfaces

## 5.1 The effect of soil nonlinearity

The final step in terms of modelling sophistication enhancement is to account for nonlinear soil response, employing the previously discussed nonlinear constitutive model. The effect of soil nonlinearity is assessed by comparing the dynamic response of the reactor building, with and without soil nonlinearity.

To focus on the role of soil nonlinearity, the auxiliary building is removed. Comparing the SA at the top of the external containment wall for the two cases (**Figure 15c**), two interesting observations can be made. First, the rocking, rigid body–like vibration mode of the reactor building with a frequency close to 2 Hz disappears with soil nonlinearity. Second, there is a remarkable increase in the primary bending vibration mode of the reactor building at around 4 Hz. The middle part of the figure (**Figure 15b**), which includes the 1.5-2 Hz and 3-4 Hz filtered time history plots, and the corresponding cross-sectional views of the unfiltered displacement magnitude contours at the bottom (**Figure 15a**), further assist in understanding these observations. Apparently, the continuing rocking motion of the reactor building is powered (or even amplified) by elastic soil response: every time the foundation rotates, the elastic soil bounces back, pushing the foundation to rotate towards the opposite direction. This is not the case when soil nonlinearity is accounted for, in which case the soil yields and accumulates permanent deformations. Therefore, the rocking motion at 2 Hz is severely suppressed due to soil nonlinearity, leading at the same time to an increase of resonance at a frequency close to 4 Hz.

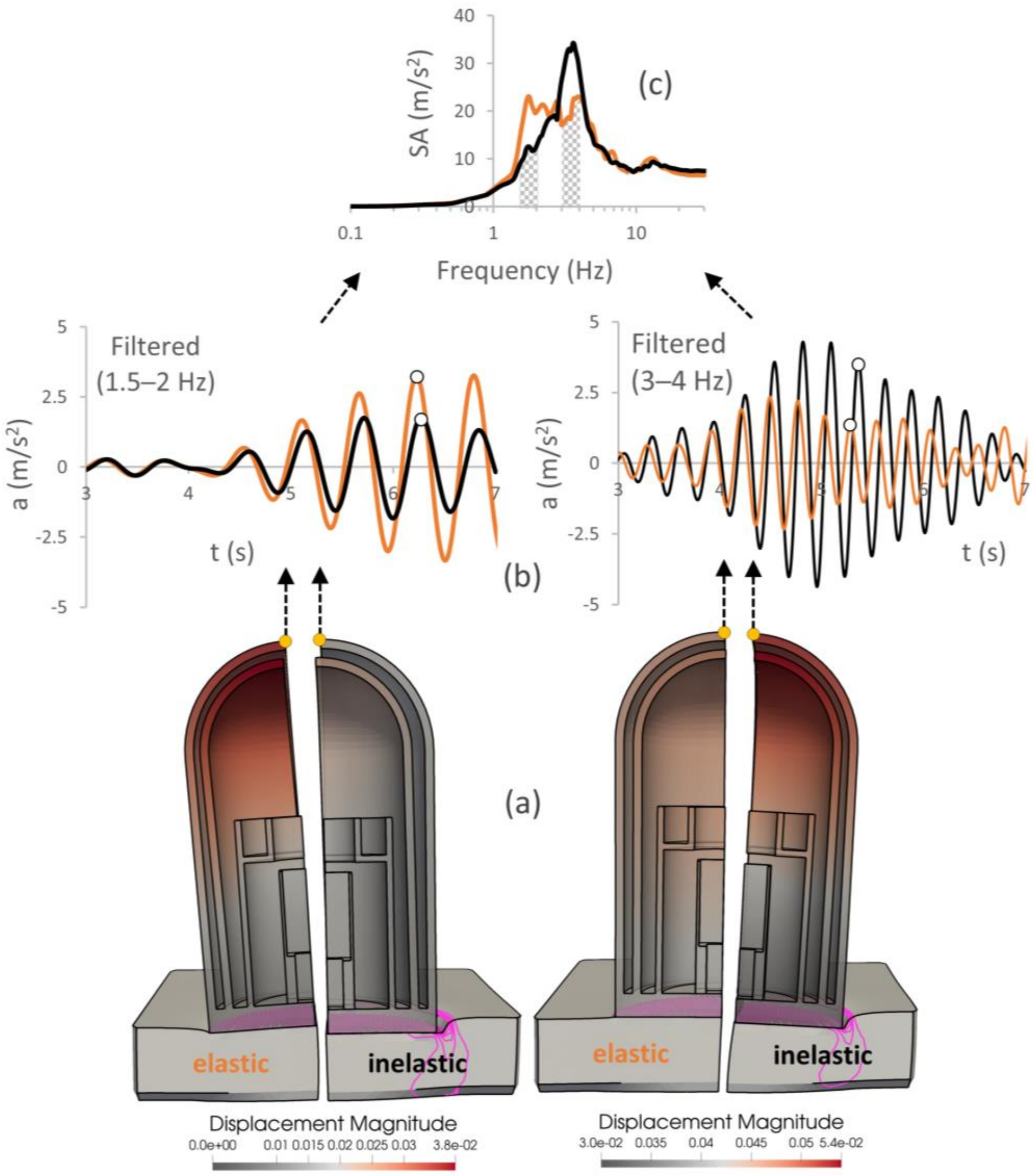

**Figure 15.** The effect of soil nonlinearity on the response of the external containment wall in the absence of the auxiliary building. Comparison with the case of elastic soil: (a) snapshots of displacement contours at *t* = 6.29 s (left) and *t* = 5.37 s (right) (scale factor 200); (b) filtered acceleration time histories at 1.5–2 Hz (left) and 3–4 Hz (right); and (c) the corresponding elastic acceleration response spectra SA.

Regarding the effect of soil nonlinearity on the response of the RV, the previously discussed amplification in the high-frequency range seems to disappear, as shown in **Figure 16d**. The cross-sectional views of the 10–13 Hz filtered displacement contours of interest at the bottom (**Figure 16a**), in combination with the time histories of foundation uplifting (**Figure 16b**), explain the beneficial role of soil nonlinearity. Evidently, soil yielding beneath the foundation leads to accumulation of permanent settlements, resulting in negligible opening and closing of the gap at the interface and dampening of any tendency for high frequency excitation. In the absence of such "gapping" mechanism, the high-frequency response of the RV cannot be excited.

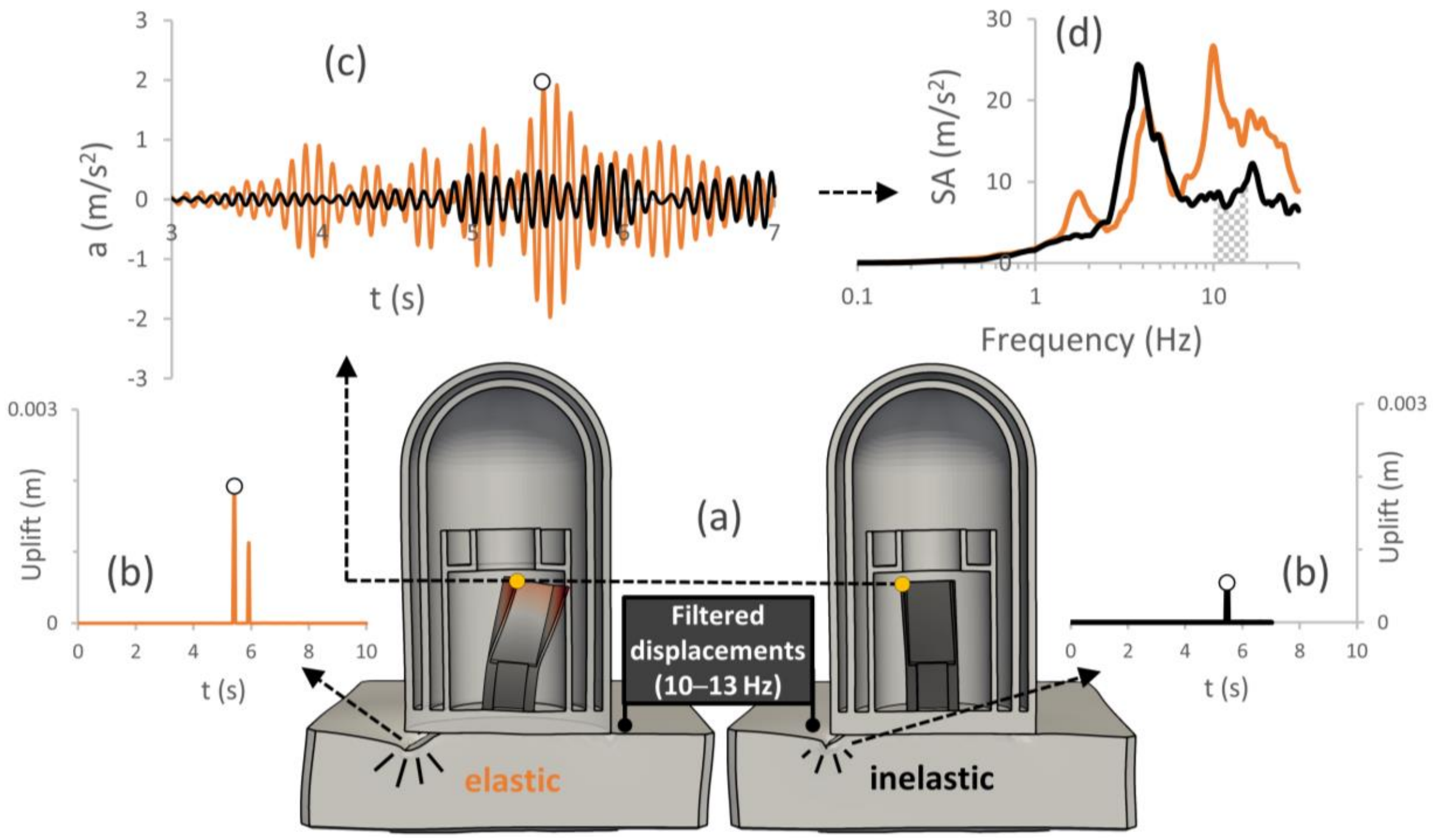

**Figure 16.** The effect of soil nonlinearity on the response of the RV in the absence of the auxiliary building. Comparison with the case of elastic soil: (a) snapshot of filtered displacement contours (10–13 Hz) at $t$ = 5.45 s (scale factor 20’000); (b) time histories of foundation uplifting; (c) filtered acceleration time histories at the top of the RV; and (d) the corresponding elastic acceleration response spectra.

To assess the influence of soil nonlinearity on the complete model, the auxiliary building is included in the analysis. **Figure 17b** compares the SA at the top of the WP (point 4) for the cases of elastic and nonlinear soil (and nonlinear interfaces for both cases), while the SA at the top of the auxiliary building (point 8) are also provided. The reduction of stiffness due to nonlinear soil response leads to a slight shift of the peak SA of point 8 to a lower frequency, while almost maintaining its magnitude. Although the response of the auxiliary building is only slightly affected by soil nonlinearity, the response of the WP is significantly altered, resulting in a significant decrease of SA. As nonlinear soil response suppresses the rocking motion of the system, the previously described horizontal out-of-phase reduction mechanism becomes more prominent, leading to a beneficial interaction between the two buildings, which is responsible for the aforementioned SA reduction. The cross-sectional views of the 3–5 Hz filtered displacement contours at the bottom (**Figure 17a**) clearly show the detrimental out-of-phase rotational interaction that prevails in the case of elastic soil, compared to the beneficial out-of-phase horizontal interaction between the two buildings that dominates when soil nonlinearity is accounted for.

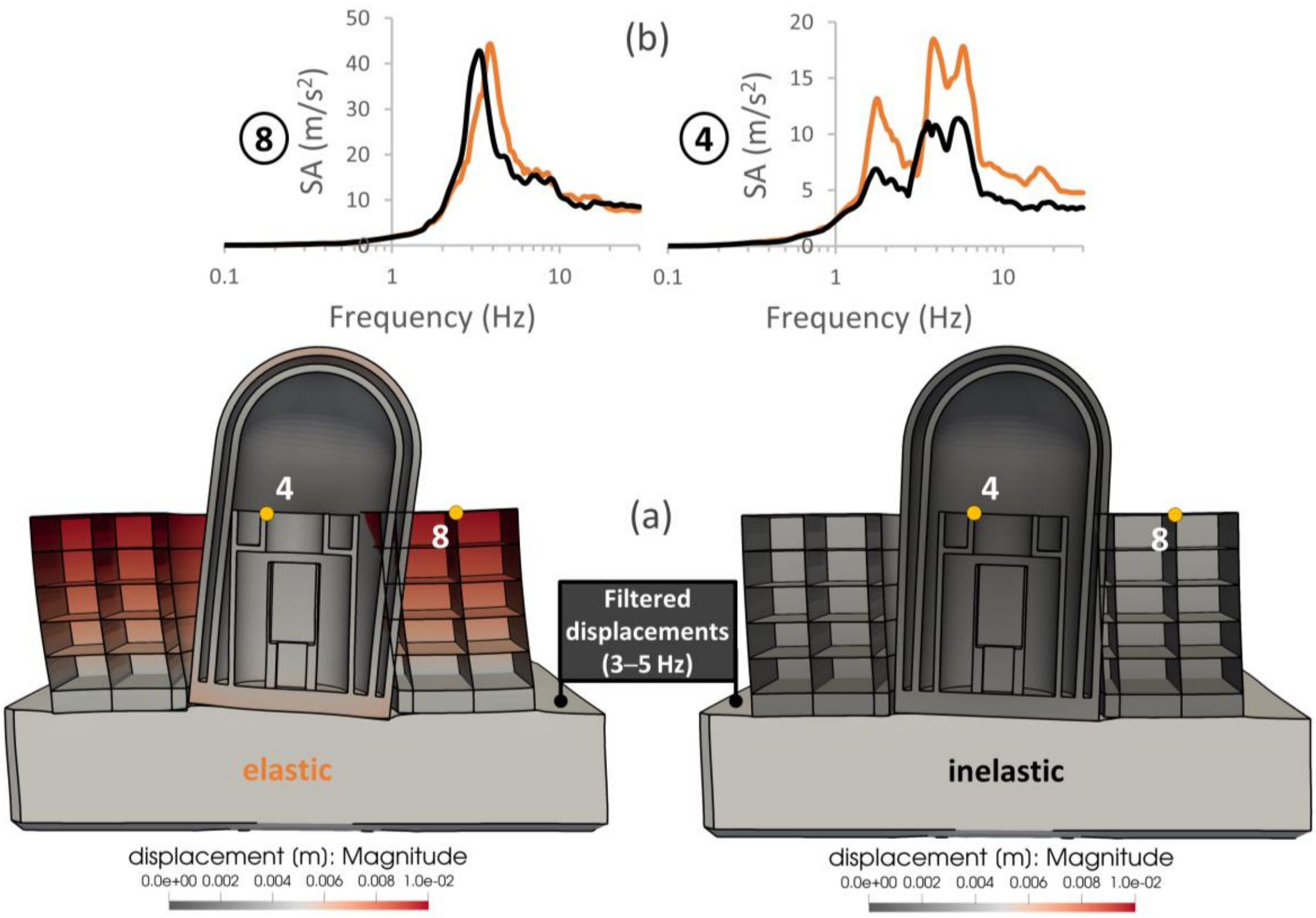

**Figure 17.** The effect of soil nonlinearity on the response of the WP in the presence of the auxiliary building. Comparison with the case of elastic soil: (a) snapshot of filtered displacement contours (3–5 Hz) at *t* = 5.72 s (scale factor 700); and (b) elastic acceleration response spectra (SA) at points 8 and 4. The suppression of the rotational vibration modes caused by soil nonlinearity results in a beneficial out-of-phase horizontal coupling between the two buildings.

## 5.2 The effect of the auxiliary building on the reactor building (SSSI)

Building on the positive effects of soil nonlinearity, which was shown to improve the dynamic interaction between the two structures, this section aims to evaluate the effect of the auxiliary building on the response of reactor building. In this context, the most sophisticated and realistic models (nonlinear soil and nonlinear interfaces) are used to compare the response of the reactor building, with and without the auxiliary building. A summary of the results is presented in **Figure 18**, allowing for several interesting observations. First, the resonant frequency of the auxiliary building (point 8) is reduced from 4 Hz in the elastic soil case to 3.3 Hz, due to soil nonlinearity. Most importantly, the presence of the auxiliary building affects the SA at all points of interest of the reactor building for frequencies higher than the 3.3 Hz resonance (see shaded areas in **Figure 18**). Evidently, the nearly shear frame-type behavior of the auxiliary building does not couple well with the external and internal containment walls of the reactor building, which are much taller such that their dynamic response is dominated by their bending vibration modes. Hence, the SA of points 5, 6, and 7 are only slightly shifted to lower frequencies due to the presence of the auxiliary building, which increases the inertia of the system. This slight decrease of the frequency is in agreement with the experimental results of Kitada et al. [51]. However, a completely different pattern is observed for the remaining points of interest, specifically for the internal CW, the dynamic characteristics (resonant frequency and height) of which are similar to those of the auxiliary building. A remarkable reduction of about 55% in SA is observed for the top of the WP (point 4). The reduction at the top of the RV (point 3) is of the same order of magnitude, while significant reductions are also observed for the remaining two points of interest.

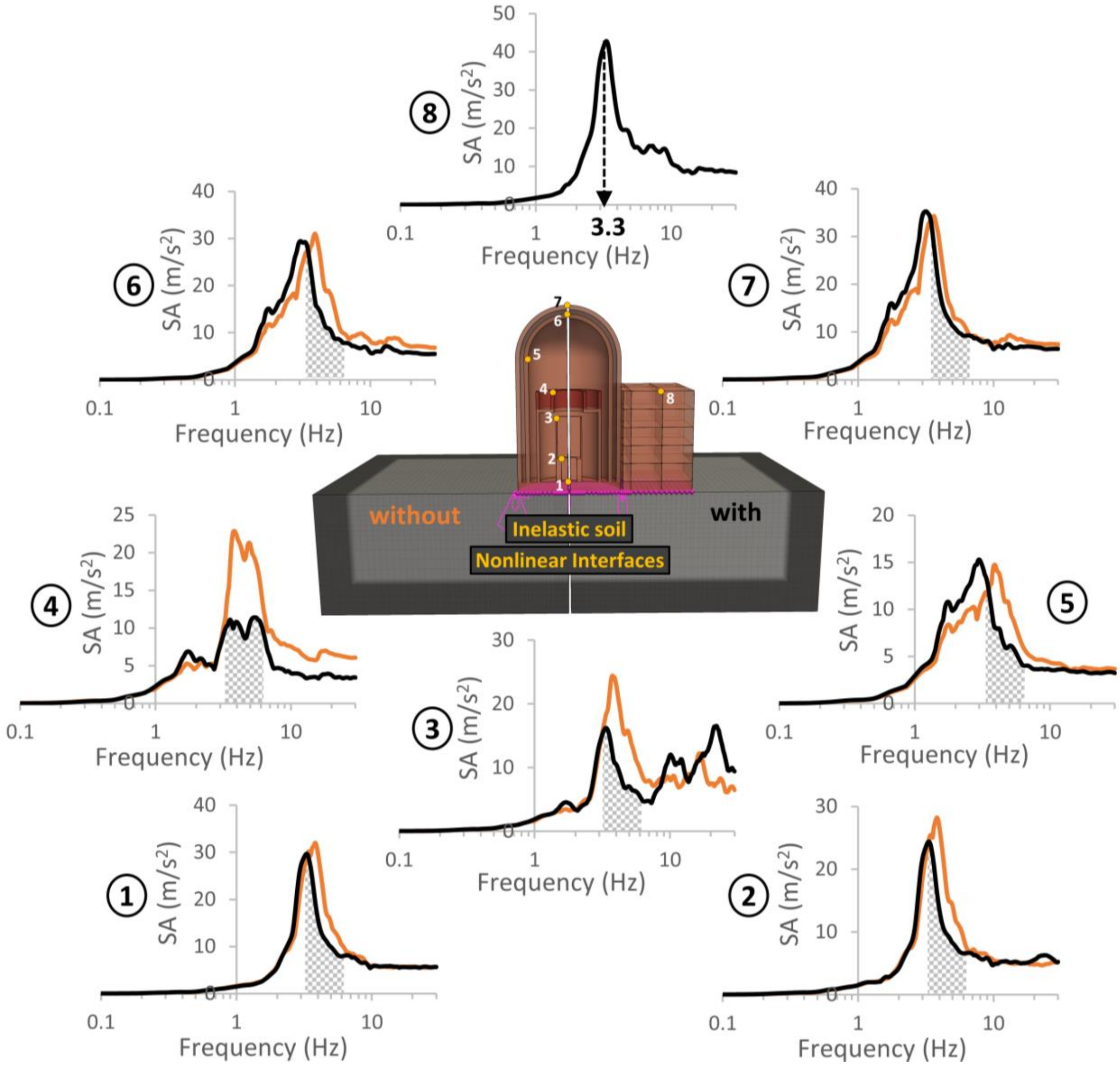

**Figure 18.** Nonlinear soil and nonlinear interfaces. The effect of SSSI on the response of the reactor building. Comparison of elastic acceleration response spectra SA at critical points of interest on the reactor building, "with" and "without" the auxiliary building.

The last two figures aim to explain the mechanism that enables the auxiliary building to act as a seismic protection device for the reactor building, similar to seismic resonant metamaterials. **Figure 19** compares the cross-sectional views of the 3–5 Hz filtered displacement contours, with and without the auxiliary building, along with the corresponding filtered acceleration time histories at the top of the WP (point 4). At the specific snapshot, the soil beneath the two structures starts moving to the left, while at the same time the auxiliary building is moving to the right. This out-of-phase oscillation of the auxiliary building relative to the soil, that occurs at frequencies higher than its resonant frequency (3.3 Hz), reduces significantly the horizontal accelerations of the soil at the 3-5 Hz frequency range. This is beneficial for the parts of the reactor building that have resonant vibration modes in this frequency range and, thus, are subjected to reduced dynamic excitation by the soil.

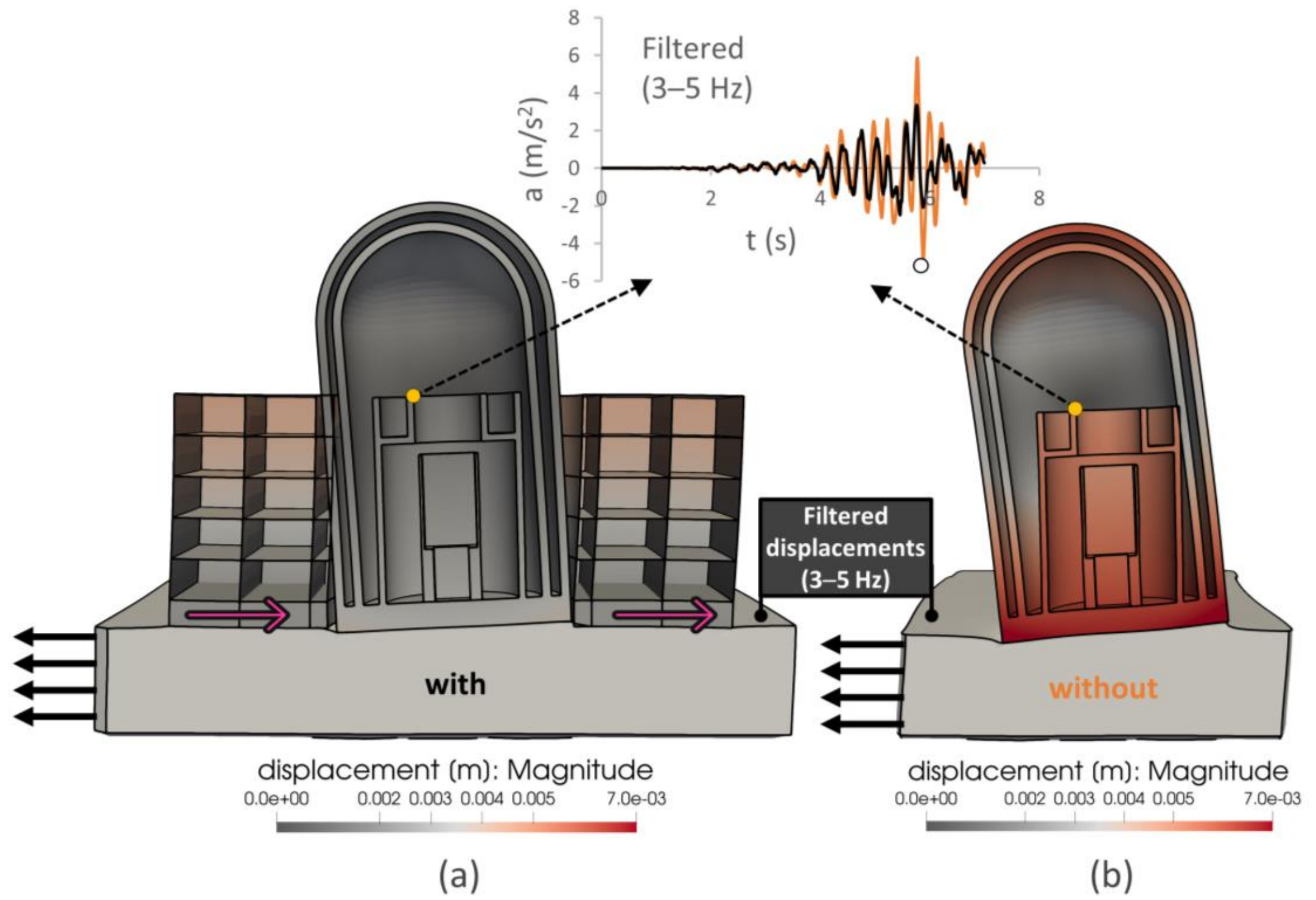

**Figure 19.** Nonlinear soil and nonlinear interfaces. Illustration of the beneficial effect of the auxiliary building to the response of the reactor building. Snapshot of filtered displacement contours (3–5 Hz): (a) "with"; and (b) "without" the auxiliary building at *t* = 5.87 s (scale factor 700). The inset plots the acceleration time histories at the top of the WP. The out-of-phase horizontal coupling between the two buildings is beneficial, reducing the excitation of the reactor building.

To visualize this out-of-phase mechanism, **Figure 20** compares the displacement vectors in the soil at the same time for the two cases examined, viewed from above. Notice that in the left snapshot (with the auxiliary building) although the soil has started moving to the left, the auxiliary building is still moving to the right pushing the soil to this direction; as a result, the horizontal displacement vectors beneath it point to the right. For the studied low-frequency range, the effect of the structures on the displacement vectors of the soil is localized, disappearing approximately at distance $B/4$ from the auxiliary building (where $B$ is the side length of the auxiliary building; **Figure 20a**), and at distance $D$ from the reactor building (where $D$ is the diameter of the reactor building; **Figure 20b**). This indicates that SSSI effects are not relevant for widely-spaced structures, but do play an important role for closely-spaced structures, such as the ones examined herein. Depending on the importance of the structures, such effects may need to be investigated in more detail. As revealed by the presented results, the effect of SSSI can be beneficial, provided that the structures are properly "tuned". In the studied idealized problem, this was a mere coincidence, but the implications of the observed *resonant metamaterial–like* response can be far reaching and are worth further consideration.

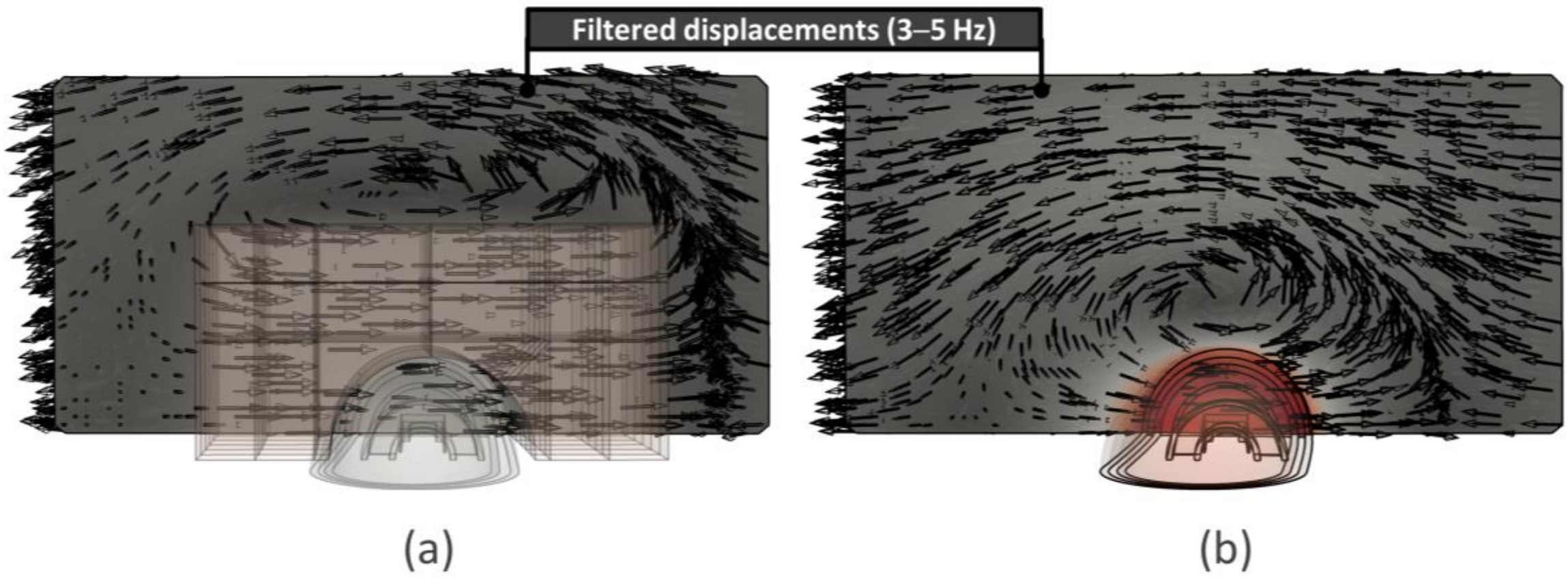

**Figure 20.** Nonlinear soil and nonlinear interfaces. Illustration of the beneficial effect of the auxiliary building to the response of the reactor building. Snapshots of filtered displacement contours (3–5 Hz): (a) “with”; and (b) “without” the auxiliary building at *t* = 5.87 s (scale factor 700). The inset displacement vectors illustrate the beneficial out-of-phase horizontal mechanism provided by the auxiliary building.

A final remark on the differential displacements between the two buildings is necessary. Since important equipment (e.g., pipes and other conduits) traverses through the reactor and the auxiliary building, there are strict differential displacement tolerances that must be met. The results presented herein are only meant to provide a crude estimation of the order of magnitude of such differential displacements. For the specific (idealized) case examined, the vertical differential settlement between the two buildings is of the order of 10 mm, while the maximum horizontal differential displacement (due to sliding at the interfaces) does not exceed 1.5 mm. Finally, the maximum horizontal differential displacement between the two buildings at the top of the auxiliary building is roughly 30 mm, indicating a potential for pounding if the buildings are placed close to each other. No such pounding was modelled in the analysis, but as previously mentioned, the buildings were spaced at 1 m distance for numerical reasons.

# 6 Conclusions

A sophisticated 3D FE model of an idealized—but based on existing designs—Nuclear Power Plant (NPP) was simulated in the Real-ESSI Simulator, aiming to investigate the dynamic interaction between the reactor building and the auxiliary building. The main objective of the study was to evaluate the effect of the auxiliary building on the seismic response of critical components of the reactor building. The effect of soil and interface nonlinearities was thoroughly investigated by gradually increasing the sophistication level of the analysis. Initially, linear elastic soil conditions were assumed and the soil–structure interfaces were tied. Then, special nonlinear interfaces were introduced, allowing for uplifting and sliding at the soil-foundation interfaces. Finally, a simple yet realistic nonlinear constitutive soil model was introduced, suitable for modeling the dynamic cyclic response of pressure-independent materials, such as clays. The seismic wave field was assumed to consist only of vertically propagating shear waves (SH), which are inserted into the model using the Domain Reduction Method (DRM), targeting a specific artificial accelerogram at the ground surface. The key findings of the study are summarized as follows:

1) Elastic soil and tied interfaces
   - In the absence of the auxiliary building, elastic SSI introduces an additional rocking–type vibration mode of the soil–reactor building system at about 2 Hz, which cannot possibly be predicted by a fixed-base model (often assumed in the past). This implies that the traditional assumption that ignoring SSI is conservative may not be true, calling for more advanced modelling that can account for such effects, especially in the case of structures of high importance such as NPPs.
   - The presence of the auxiliary building and the developing SSSI, leads to an overall amplification of spectral accelerations SA at characteristic critical points of the reactor building, at frequencies close to the resonance of the auxiliary building. This amplification is associated with a *detrimental out-of-phase* *rotational* interaction mechanism between the two buildings, which leads to an increase of the rotational response of the reactor building.

2) Elastic soil and nonlinear interfaces
   - The response is significantly affected by the nonlinear interfaces. During the rocking oscillation of the reactor building, the gapping mechanism (opening and closing) at the soil–foundation interface leads to the development of additional higher-frequency excitations. This unavoidably leads to significant amplifications of response above 10 Hz, which can be *detrimental* for internal components, such as the Reactor Vessel (RV).
   - The previously mentioned *detrimental* effect of the auxiliary building on the reactor building due to their out-of-phase rotational interaction is slightly reduced due to limited sliding at the soil-foundation interface of the auxiliary building, which dissipates energy and slightly reduces the response of the auxiliary building.
3) Nonlinear soil and nonlinear interfaces
   - Nonlinear soil response leads to a significant suppression of the rocking vibration mode of the reactor building. This completely changes the *detrimental out-of-phase rotational interaction* of the two buildings, which now becomes a ***beneficial out-of-phase horizontal interaction*** for frequencies near and above the resonance of the auxiliary building.
   - Resembling the seismic protection mechanism of *seismic resonant metamaterials*, the auxiliary building essentially protects the reactor building by moving out-of-phase (at its resonant frequency and higher) relative to the soil, thus reducing the excitation of the latter, and consequently the response of critical components inside the reactor building.
   - The components that benefit the most from such *beneficial out-of-phase interaction* between the two buildings are the Cylindrical Wall (CW) and the RV, which experience a remarkable reduction in spectral accelerations SA of the order of 55%, in the frequency range associated with the resonant frequency of the auxiliary building.

Summarizing, the present study has highlighted the importance of accounting for soil and interface nonlinearities, while showing that SSI and SSSI can be beneficial or detrimental for the response of the structure, and their effects *cannot* be known *a priori*. The *beneficial* effects of SSSI between the reactor and the auxiliary building, acting in a similar manner as *resonant metamaterials,* underlines the importance of accounting for such effects in the design of new and the evaluation of existing structures of critical importance, such as NPPs. Optimizing the dynamic characteristics of neighboring structures, aiming to maximize their beneficial SSSI may lead to more earthquake-resilient clusters of structures, or even entire *Meta-cities*, featuring tuned ensembles of building clusters, optimized to reduce the combined seismic risk.

# 7 Acknowledgements

This research was part of the project 'Inspire' funded by the European Union's Horizon 2020 research and innovation programme under the Marie Skłodowska-Curie grant agreement no. 813424. Additional support from ETH Zurich and University of California Davis is gratefully acknowledged.

# 8 Data availability statement

The input files, as well as the pre-and-postprocessing python scripts developed for this paper, have been archived in the ETH Research Collection [52] and ETH Data Archive Deposit (only the scripts) ([53,54]). The latest versions of these python scripts are also available on GitHub (https://github.com/ConstantinosKanellopoulos).